\documentclass[%
 reprint,
superscriptaddress,
preprintnumbers,
 amsmath,amssymb,
 aps,
 prl,
]{revtex4-2}

\usepackage{graphicx}
\usepackage{dcolumn}
\usepackage{bm}
\usepackage{xcolor}

\usepackage[normalem]{ulem} 
\newcommand{\stkout}[1]{\ifmmode\text{\sout{\ensuremath{#1}}}\else\sout{#1}\fi}

\usepackage{amsthm}

\newtheorem*{theorem}{Theorem}

\begin{document}

\title{Estimation of Thermodynamic Observables in Lattice Field Theories with Deep Generative Models}

\author{Kim A. Nicoli}
\affiliation{Machine Learning Group, Technische Universit\"{a}t Berlin, Berlin, Germany}

\author{Christopher J. Anders}
\affiliation{Machine Learning Group, Technische Universit\"{a}t Berlin, Berlin, Germany}

\author{Lena Funcke}
\affiliation{Perimeter Institute for Theoretical Physics, Waterloo, Canada}

\author{Tobias Hartung}
\affiliation{Department of Mathematics, King’s College London, London, United Kingdom}

\author{Karl Jansen}
\affiliation{NIC, DESY Zeuthen, Zeuthen, Germany}

\author{Pan Kessel}
\affiliation{Machine Learning Group, Technische Universit\"{a}t Berlin, Berlin, Germany}

\author{Shinichi Nakajima}
\affiliation{Machine Learning Group, Technische Universit\"{a}t Berlin, Berlin, Germany}
\affiliation{RIKEN Center for AIP, Tokyo, Japan}

\author{Paolo Stornati}
\affiliation{NIC, DESY Zeuthen, Zeuthen, Germany}
\affiliation{Institut f\"{u}r Physik, Humboldt-Universit\"{a}t, Berlin, Germany}

\preprint{HU-EP-20/16}

\begin{abstract}
In this work, we demonstrate that applying deep generative machine learning models for lattice field theory  
is a promising route for solving problems where Markov Chain Monte Carlo (MCMC) methods are problematic. 
More specifically, we show that generative models can be used to estimate the absolute value of the 
free energy, which is in contrast to existing MCMC-based methods which are limited to only 
estimate free energy differences. We demonstrate the effectiveness of the proposed method for two-dimensional $\phi^4$ theory and compare it to MCMC-based methods in detailed numerical experiments.
\end{abstract}

\maketitle

\paragraph{Introduction.}
The free energy of a physical system is of great importance since it can be related to several thermodynamical observables. In particular, at non-zero temperature, it allows to compute the entropy, the pressure or, more generally, the equation of state of the considered physical system. 
For example, QCD at high temperature --as a generic strongly interacting field theory-- plays an essential role in the physics of the early universe and is now extensively probed in large-scale heavy ion experiments \cite{busza2018heavy}. Hence, knowing such thermodynamic quantities from QCD alone is of very high relevance.

The main tool to study strongly-coupled field theories, such as QCD, is to discretize 
them on a spacetime lattice and use Monte-Carlo Markov-Chain (MCMC) methods to numerically 
calculate the relevant physical quantities. Unfortunately, these thermodynamical quantities are challenging to compute using existing MCMC methods. The fundamental difficulty is that MCMC is not able to directly estimate the partition function of the lattice field theory. Therefore, the absolute value of the free energy cannot be estimated straightforwardly.

Instead, there are a number of MCMC methods to estimate differences of free energies. One typically chooses a free energy difference $\Delta F = F_b - F_a$ such that $F_a$ is known either exactly or approximately. One can then deduce the value of the free energy $F_b = \Delta F + F_a$ at the desired point in parameter space. If the free energy $F_a$ is not known exactly, this induces an unwanted approximation error.
Most of the methods to estimate $\Delta F$ rely on integrating a derivative of the partition function over a trajectory in the parameter space of the lattice field theory \cite{philipsen2013qcd}. Alternatively, one can use a reweighting procedure to calculate free energy differences between neighbouring points of the discretized trajectory and then sum them up \cite{philipsen2013qcd, de2001t}. These approaches require simulations at each parameter point of the discretized trajectory which is numerically costly and leads to accumulation of errors. This effect is often the dominant contribution to the error - especially if the trajectory passes a phase transition. Such situations arise for example in the context of studying the deconfined phase of $SU(3)$ Yang--Mills theory \cite{giusti2017equation, caselle2018qcd}. We stress that the accumulation of the statistical error along the trajectory and the approximation error of its starting point are not independent. The former could be reduced if a better starting point was available. There are also non-equilibrium methods based on Jarzynski’s identity to estimate free energy differences without the need for integration \cite{jarzynski1997equilibrium, jarzynski1997nonequilibrium, caselle2018qcd}. However, also these methods require expensive repeated simulations corresponding to an ensemble of non-equilibrium trajectories through phase-space.

It is therefore desirable to develop methods which allow the direct estimation of the free energy at a given point in parameter space. 

In the following, we will propose such a method based on deep generative machine learning models. As we will discuss, our method comes with rigorous error estimators and asymptotic guarantees. Over the last years, deep generative models have been applied with great success to generate, for example, high-resolution images, natural speech, and text (see \cite{goodfellow2016deep} for an overview). In \cite{shanahan2018machine}, a machine-learning-based regression algorithm for determining action parameters from an ensemble of field configurations is proposed and \cite{bachtis2020mapping} uses a neural network to predict the structure of phase transitions from field configurations. References \cite{hol1, hol2, hol3, hol4, hol5} conjecture a relation between Restricted Boltzmann Machines and Quantum Fields in the context of the holographic duality.
In the recent works \cite{flowsforlattice, flowsforlattice2, flowsforlattice3}, deep generative models have also been used in the context of lattice quantum field theories (see also \cite{urban2018reducing,tanaka2017towards}). The main objective of these works was to reduce the integrated autocorrelation of the simulations. In contrast, this work demonstrates that deep generative models can be used to estimate quantities which are not (directly) obtainable by MCMC approaches. 

We also note that generative models have been used in \cite{wirnsberger2020targeted} to estimate free energy differences in the context of statistical mechanics by combining these models with the Zwanzig free energy perturbation method \cite{zwanzig1955high}. Contrary to this approach, our method estimates the absolute value of the free energy. We furthermore note that the free energy can also be directly computed using the Tensor Renormalization Group method, see \cite{akiyama2020tensor} for an application to $\phi^4$-theory. For other novel approaches to obtain thermodynamic quantities and, in particular, the equation of state, see \cite{asakawa2014thermodynamics, giusti2014equation}.

In the following, we will give a brief overview of relevant aspects of lattice field theories and generative models. We will then discuss how generative models can be used to estimate the free energy and compare this approach to MCMC-based methods in numerical experiments.

\paragraph{Lattice Field Theory.}

A lattice field theory can be described by an action $S(\phi)$. In the following, we will consider (euclidian) real scalar field theory for concreteness, i.e. $\phi(x) \in \mathbb{R}$ for each lattice site $x \in \Lambda$ of the lattice $\Lambda$. The path integral then reduces to an ordinary high-dimensional integral. Therefore, expectation values of operators $\mathcal{O}(\phi)$ can be calculated by
\begin{align*}
    \langle \mathcal{O} \rangle = \frac{1}{Z} \, \int \mathcal{D}[\phi]\, \mathcal{O}(\phi)  \exp(-S(\phi)) \,, 
\end{align*}
where we defined $\mathcal{D}[\phi]= \prod_{x \in \Lambda} \textrm{d}[\phi(x)]$ and the partition function $Z$ is given by
\begin{align*}
    Z = \int \mathcal{D}[\phi]  \, \exp(-S(\phi)) \,.
\end{align*}
If we impose periodic boundary conditions in time for a lattice with temporal extend $N_T$, the theory is at finite temperature $T = \frac{1}{\beta} = \frac{1}{N_T a}$, where $a$ denotes the lattice spacing.
The free energy is then defined by
\begin{align}
    F = - T \, \ln (Z) \,,
\end{align}
and can be related to the pressure $p = - \frac{F}{V}$, where $V$ denotes the spatial volume of the lattice $\Lambda$ whose number of lattice sites we denote by $|\Lambda|$. Similarly, the entropy $H$ can be obtained from the free energy by $F = U - T H$, where the $U$ is the internal energy.

\paragraph{Deep Generative Models.}

We focus on a particular subclass of generative models called normalizing flows (see \cite{nfreview} for a recent review). These flows are distributions $q_\theta$ with learnable parameters $\theta$. They also have the appealing property that they allow for efficient sampling and calculation of the probability of the samples.

In more detail, these flows are constructed by defining an invertible neural network $g_\theta$. For a brief overview of neural networks, we refer to the Supplement. The samples $\phi \in \mathbb{R}^{|\Lambda|}$ are obtained by applying this network to samples $z \in \mathbb{R}^{|\Lambda|}$ drawn from a simple prior distribution $q_Z$ such as a standard normal $\mathcal{N}(0,1)$:
\begin{align}
    \phi = g_\theta(z) \,, && z \sim q_Z \,. \label{eq:nfsampling}
\end{align}
Since the network $g_\theta$ is invertible by assumption, it then follows by the change of variable theorem that $\phi \sim q_\theta$ with
\begin{align}
    q_\theta(\phi) &= q_Z(g^{-1}_\theta(\phi)) \, \left| \frac{\textrm{d}g_\theta}{\textrm{d}z} \right|^{-1} \,. \label{eq:nfprob}
\end{align}
The architecture of the neural network $g_\theta$ is chosen such that i.) invertibility of $g_\theta$ and ii.) efficient evaluation of the Jacobian determinant $\left| \tfrac{\textrm{d}g_\theta}{\textrm{d}z} \right|$ are ensured. A particular example of such an architecture is \emph{Non-linear Independent Component Estimation} (NICE) \cite{nice} for which the neural network $g_\theta$ consists of invertible coupling layers $y^l:\mathbb{R}^{|\Lambda|} \to \mathbb{R}^{|\Lambda|}$, i.e.
\begin{align}
    g_\theta(z) = \left(y^L \circ y^{L-1} \circ \dots \circ y^1 \right) (z) \label{eq:g}
\end{align}
Invertiblity and efficient evaluation of Jacobian determinant is then ensured by splitting the components of the layer $y^l = (y^l_u, y^l_d)$ in two parts $y^l_u \in \mathbb{R}^{|\Lambda|-k}$ and $y^l_d \in \mathbb{R}^{k}$ for given $k\in \{1, |\Lambda| -1 \}$. The layer $y^{l+1} = (y^{l+1}_u, y^{l+1}_d)$ is then recursively defined by
\begin{align}
y^{l+1}_u &= y^l_u \,, \nonumber \\ 
y^{l+1}_d &= y^l_d + m( y^l_u ) \label{eq:couplinglayer} \,,
\end{align}
where $m$ is another neural network (not necessarily satisfying the two requirements from above). 
Due to the splitting, this can be easily inverted by
\begin{align*}
y^{l}_u &= y^{l+1}_u \,, \\
y^{l}_d &= y^{l+1}_d - m( y^{l+1}_u ) \,,
 \end{align*}
and the determinant of the Jacobian is given by
\begin{align*}
\det \frac{\partial y^{l+1}}{\partial y^l} =
\begin{vmatrix}
\frac{\partial y^{l+1}_u}{\partial y^{l}_u} & \frac{\partial y^{l+1}_u}{\partial y^{l}_d}  \\
\frac{\partial y^{l+1}_d}{\partial y^{l}_u} & \frac{\partial y^{l+1}_d}{\partial y^{l}_d} \\
\end{vmatrix}
=
\begin{vmatrix}
\mathbb{I} & 0 \\
* & \mathbb{I} \\
\end{vmatrix}
= 1 \,. 
\end{align*}
The total Jacobian determinant is then $\left| \frac{\textrm{d}g_\theta}{\textrm{d}z} \right|=1$ since it is the product of the Jacobian determinant of each layer.

\paragraph{Training.}

We want to train a generative model which samples field configurations $\phi \sim q_\theta$ approximately from the path-integral distribution
\begin{align}
    p(\phi) = \frac{1}{Z} \, \exp(-S(\phi)) \,. \label{eq:targetdist}
\end{align}
For this, the Kullback--Leibler divergence \cite{mackay2003information} between the normalizing flow $q_\theta$ and the target distribution $p$ is minimized, i.e.
\begin{align*}
    \textrm{KL}(q_\theta || p) &= \int \mathcal{D}[\phi]\, q_\theta(\phi) \, \ln \left( \frac{q_\theta(\phi)}{p(\phi)} \right) \\
    &= \beta \left( F_q - F \right) \,,
\end{align*}
where we have defined the variational free energy 
\begin{align}
    \beta F_q  &= \mathbb{E}_{\phi \sim q_\theta} \left[ S(\phi) + \ln q_\theta(\phi) \label{eq:varfreeenergy} \right] 
\end{align}
as well as the expectation value $\mathbb{E}_q[\mathcal{O}]=\int \mathcal{D}[\phi] q_\theta(\phi) \mathcal{O}(\phi)$ and the free energy $F=-\tfrac{1}{\beta} \ln(Z)$. This divergence vanishes if and only if the distributions $q$ and $p$ are identical \footnote{More precisely, the distributions $p$ and $q$ have to be identically only almost everywhere, i.e. up to a set of measure zero.}.

The KL divergence is minimized by gradient descent with respect to the parameter $\theta$ of the flow $q_\theta$. Since the free energy $F$ does not depend on the flow $q$, the variational free energy $F_q$ can equivalently be minimized. Therefore, the training procedure does not require a target distribution \eqref{eq:targetdist} with a tractable partition function $Z$. 
Using the explicit expression for the probability of the flow \eqref{eq:nfprob}, we can rewrite the variational free energy as
\begin{align*}
    \beta F_q 
    &= \mathbb{E}_{z\sim q_Z} \left[ S(g_\theta(z)) - \ln \left| \frac{\textrm{d}g_\theta}{\textrm{d}z} \right|(z)  + \ln q_Z(z) \right] \,.
\end{align*}
In training, the expectation value is approximated by its Monte-Carlo estimate.
In machine learning, this approach of learning a model from an unnormalized target distribution is very well established \cite{kingma2013auto, rezende2014stochastic, rezende2015variational, muller2019neural}. Recently, the same method has been used in the context of lattice field theories \cite{flowsforlattice}. Furthermore, this approach has been applied to quantum chemistry \cite{noe2019boltzmann} and statistical physics \cite{wu2019solving, nicoli2020asymptotically, nicoli2019comment}. 

The variational free energy does not allow us to infer the value of the KL divergence since the free energy $F$ is not known. In order to alleviate this shortcoming, we define the random variable $ C(\phi) = S(\phi) + \ln q_\theta(\phi)$, which is related to the variational free energy by $\beta F_q = \langle C \rangle_q$. In the Supplement, we show that
\begin{align*}
    \textrm{KL} (q_\theta || p) = \tfrac{1}{2} \, \textrm{Var}_q(C) +  \mathcal{O}(\mathbb{E}_q [|w - 1|^3]) \,,
\end{align*}
where we have defined the importance weight $w(\phi) = \tfrac{p(\phi)}{q(\phi)}$. 
Thus convergence of training will result in a small variance $\textrm{Var}_q(C)$. In practice, a Monte-Carlo estimate of this quantity can be calculated without any significant overhead during training as $C(\phi)$ is also needed for Monte-Carlo estimation of the variational free energy $F_q$, see \eqref{eq:varfreeenergy}. It is therefore advisable to closely monitor the variance of $C$ during training.

\begin{figure}[tb]
    \centering
    \includegraphics[width=0.45\textwidth]{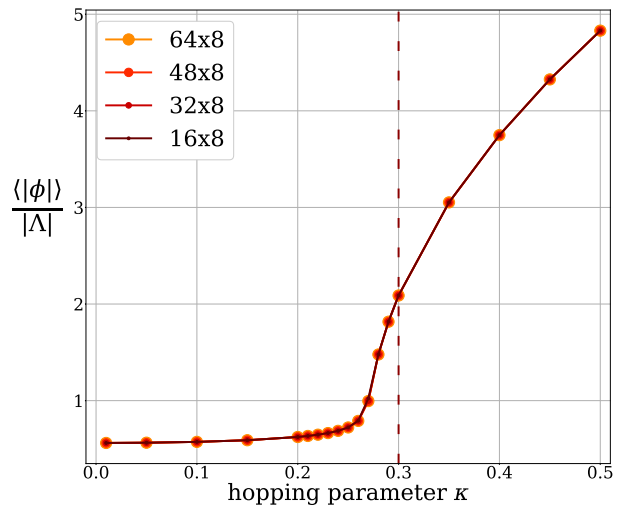}
    \caption{Absolute magnetization density as a function of hopping parameter $\kappa$ for bare coupling $\lambda=0.022$. Results for various lattice sizes overlap. The values were estimated with an overrelaxed HMC \cite{adler1981over, whitmer1984over, callaway1983lattice, fodor1994overrelaxation}. The dashed line denotes the hopping parameter value $\kappa=0.3$ for the free energy estimation in the numerical experiments.}
    \label{fig:absmag}
\end{figure}

\begin{figure}[tb]
    \centering
    \includegraphics[width=0.45\textwidth]{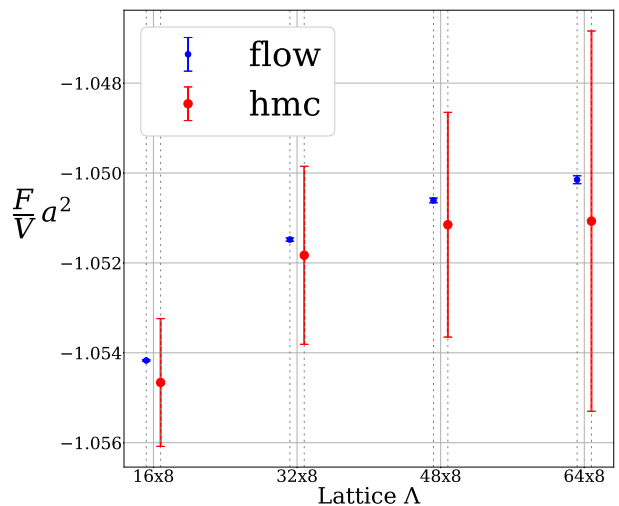}
    \caption{Estimate of free energy density at $\lambda=0.022$ and $\kappa=0.3$ obtained by both the \textcolor{blue}{\textbf{flow-based}} and \textcolor{red}{\textbf{MCMC-based}} method for various lattice sizes. MCMC estimates are obtained from integrating free energy differences. Both methods use the same number of samples (5.6 M) for estimation. Errors are obtained with the delta and uwerror method \cite{wolff2004monte} for flow and HMC respectively (see Appendix for Jackknife error analysis).}
    \label{fig:freeeneergy}
\end{figure}
\paragraph{Estimation of Thermodynamical Observables.}\label{par:therm_observables}

The partition function $Z$ can be rewritten as
\begin{align}
    Z = \int \mathcal{D}[\phi] \, q_\theta(\phi) \, \tilde{w}(\phi) \,, \label{eq:impweightZ} 
\end{align}
where we have defined the unnormalized importance weight $\tilde{w}(\phi)=\frac{\exp(-S(\phi))}{q_\theta(\phi)}$. Therefore, the partition function can be estimated by Monte-Carlo as follows
\begin{align}
    \hat{Z} = \frac{1}{N} \sum_{i=1}^N \tilde{w} (\phi_i) && \textrm{with} && \phi_i \sim q_\theta \,. \label{eq:partition_est}
\end{align}
We emphasize that the sampling procedure does not need to be sequential (as for a Markov Chain). As a result, it can very efficiently be parallelized and does not suffer from autocorrelation.
From $\hat{Z}$, one can then easily estimate the free energy by
\begin{align}
    \hat{F} = - T \, \ln \hat{Z} \,. \label{eq:freeenergyest}
\end{align}
From the free energy \eqref{eq:freeenergyest}, one can then straightforwardly obtain estimates for the pressure and entropy, as explained above.
The estimator \eqref{eq:freeenergyest} has been extensively studied in the context of training an Importance Weighted Variational Autoencoder (IWAE) \cite{burda2015importance, nowozin2018debiasing, teh2007collapsed}. It was shown in \cite{burda2015importance} that it is a statistically consistent estimator if $q$ has support larger or equal to the target $p$. In \cite{nowozin2018debiasing}, its variance and bias were derived using the delta method (see also \cite{teh2007collapsed, nicoli2020asymptotically}). For convenience, we summarize the relevant results in the Supplement. Alternatively, one can use the Jackknife method to estimate the bias and variance \cite{gattringer2009quantum}.

\paragraph{Numerical Experiments.}

We apply the proposed method to two-dimensional real scalar field theory with action
\begin{align*}
S = \sum_{x \in \Lambda} - 2 \kappa \sum_{\hat{\mu}=1}^2 \varphi(x) \varphi(x + \hat{\mu}) + (1 - 2 \lambda)& \varphi(x)^2 \\
+ &\lambda \, \varphi(x)^4 \,, \label{eq:action}
\end{align*}
where $\kappa$ is the hopping parameter and $\lambda$ denotes the bare coupling constant of the theory. The action is invariant under $\mathbb{Z}_2$-transformations, i.e. $\phi \to - \phi$. Figure~\ref{fig:absmag} shows the absolute magnetization $\langle | \phi | \rangle$ as a function of the hopping parameter $\kappa$. As the hopping parameter $\kappa$ increases, spontaneous magnetization is observed.  

In the following, we will estimate the free energy $F_e$ at $\lambda_e=0.022$ and $\kappa_e=0.3$ for lattice sizes $|\Lambda| = N_L\times N_T$ of $64 \times 8$, $48 \times 8$, $32 \times 8$, $16 \times 8$ with both the flow-based and an MCMC-based method. 

Using the flow method, we can directly estimate these free energies. We modify the NICE architecture to ensure that the flow $q_\theta$ is invariant under $\mathbb{Z}_2$-transformations, i.e.
$q_\theta(\phi) = q_\theta(-\phi)$.
By the definition \eqref{eq:nfprob} of $q_\theta$, an odd function $g_\theta(-z) = -g_\theta(z)$ implies $\mathbb{Z}_2$-invariance of $q_\theta$. The map $g_\theta$ is odd if all its coupling blocks $y^l$ are odd, see \eqref{eq:g}. The latter condition can be ensured by choosing an odd neural network $m$ for the coupling \eqref{eq:couplinglayer} which we achieve by using $\textrm{tanh}$ non-linearities and vanishing biases for the network $m$. 

After training has completed, the free energy is then computed using the proposed estimator \eqref{eq:freeenergyest}. For error analysis, we use both the Jackknife as well as the delta-method and check that they lead to consistent error estimates.
In many applications, generative models suffer from \emph{mode dropping} \cite{bishop2006pattern}, i.e. some modes of the target $p$ are not captured by the model $q_\theta$. For our specific estimation method however, a simple consistency check can be performed ensuring that mode dropping does not occur. To this end, we estimate 
$Z = (\mathbb{E}_p [ \tilde{w}^{-1}])^{-1}$ by a single Markov chain at the target point in parameter space and ensure that this leads to a compatible estimate, see Supplement.

For MCMC, we use a reweighting procedure \cite{de2001t,philipsen2013qcd} which is significantly more involved and uses the relation $F_e = \Delta F_{e \, b}  + F_b$. Here, $F_b$ is the free energy at $\kappa_b = 0$ and $\lambda_b = \lambda_e$. The value of $F_b$ can be analytically calculated since for vanishing Hopping parameter $\kappa$: 
\begin{align*}
    F(\lambda) = - |\Lambda| \, T \, \ln z(\lambda) \,, 
\end{align*}
where $|\Lambda|$ denotes the number of sites of the lattice $\Lambda$ and
\begin{align*}
z(\lambda) = \sqrt{\frac{1-2\lambda}{4 \lambda}} \, \exp\left(\frac{(1-2 \lambda)^2}{8 \lambda}\right) \, K_{\frac{1}{4}}\left( \frac{(1-2 \lambda)^2}{8 \lambda} \right) \,,
\end{align*}
with $K_{n}$ being the Bessel function of the second kind. We prove this relation in the Supplement. 
The free energy difference $\Delta F_{e\,b} = F_e - F_b = - T \ln \frac{Z_e}{Z_b}$ can be obtained by
\begin{align*}
   \mathbb{E}_{p_b} \left[ \frac{\exp(-S_e)}{\exp(-S_b)} \right]= \frac{1}{Z_b} \int \mathcal{D}[\phi] \, e^{-S_b(\phi)} \,\frac{e^{-S_e(\phi)}}{e^{-S_b(\phi)}} = \frac{Z_e}{Z_b} \,.
\end{align*}
We estimate this expectation value with an overrelaxed HMC algorithm \cite{adler1981over, whitmer1984over, callaway1983lattice, fodor1994overrelaxation}.
In practice, the variance of the estimator will become prohibitively large if the two distributions $p_b$ and $p_e$ do not have sufficient overlap. We therefore choose intermediate distributions $p_{i_1},\dots p_{i_K}$ ensuring that neighbouring distributions $p_{i_k}$ and $p_{i_{k+1}}$ have sufficient overlap. The free energy difference can then be obtained by
\begin{align*}
    \Delta F_{e \, b} = \Delta F_{e, i_K} + \Delta F_{i_K \, i_{K-1}} + \dots + \Delta F_{i_1 \, b} \,.
\end{align*}
In our numerical experiments, we keep $\lambda=0.022$ fixed and only vary the hopping parameter $\kappa$ of the intermediate distributions $p_i$. We choose a difference in hopping parameter of $\delta \kappa=0.01$ for $\kappa \in [0.2, 0.3]$ and $\delta \kappa=0.05$ for all other intermediate hopping parameters $\kappa$. We therefore use $K=14$ Markov chains with 400k steps each. Thus, a total number of 5.6 million configurations is used for estimation. For a detailed analysis of the dependence of our results on this choice of $\delta \kappa$, we refer to the Supplement.

The error analysis is performed with both the uwerr \cite{wolff2004monte} and Jackknife method which are checked to lead to consistent estimates. 
We again refer to the Supplement for a more detailed description.

Figure~\ref{fig:freeeneergy} shows that the estimates of both the flow and MCMC are compatible within errorbars. However, the trajectory of the MCMC method has to pass the critical region which is challenging due to critical slowing down. The flow-based estimate can be directly performed at the desired point in parameter space and therefore does not suffer from this problem. This conceptual difference leads to a significantly more precise estimate by the flow-based method. For regions in parameter space which do not require the crossing of a phase transition, MCMC-based methods have errors of comparable order of magnitude (see Supplement). The ability of the flow to perform direct estimates is both of practical as well as of conceptual importance. For example in finite-temperature QCD, one often uses a trajectory whose initial free energy is approximated by the Hadron Resonance model (see for example \cite{philipsen2013qcd}) leading to an undesirable systematic error. Furthermore, summing up the free energy differences along the trajectory leads to an accumulation of errors.  This effect is often the dominant contribution to the error and is particularly pronounced in situations for which the trajectory has to cross a phase transition. Such situations are of great practical relevance, for example in the deconfined phase of $SU(3)$ gauge theory \cite{giusti2017equation, caselle2018qcd}. We stress that both error sources are related since the initial free energy is the starting point of the trajectory.

\paragraph{Conclusion.}
In this letter, we have proposed a method to directly estimate the free energy and hence thermodynamical observables of lattice field theories using deep generative models. This method is of great conceptual appeal as it avoids cumbersome integration through parameter space and does not require an exactly or approximately known integration constant. Future work will focus on scaling this approach to four-dimensional gauge theories. Recent work has successfully constructed flows which are manifestly gauge-invariant \cite{flowsforlattice2, flowsforlattice3}. This recent progress, combined with the enormous ongoing advances in deep learning, makes it very promising that our method can be applied to non-abelian gauge theories, and ultimately QCD, in the not too distant future.\\

\paragraph{Acknowledgements.}
This work was supported in part by the German Ministry for Education and Research (BMBF) under Grants 01IS14013A-E, 01GQ1115, 01GQ0850, 01IS18025A and 01IS18037A. This work is also supported by the Fraunhofer Heinrich-Hertz-Institut (HHI) and by the grant funded by the DFG (EXC 2046/1, Project-ID 390685689). P.S. thanks the Helmholtz Einstein International Berlin Research School in Data Science (HEIBRiDS) for funding. Research at Perimeter Institute is supported by the Government of Canada through the Department of Innovation, Science and Economic Development Canada and by the Province of Ontario through the Ministry of Economic Development, Job Creation and Trade. We want to express our gratitude for valuable feedback by Klaus-Robert M\"uller, Wojciech Samek, Nils Strodthoff, Alessandro Nada, and Stefan K\"uhn.


\bibliography{references}

\providecommand{\noopsort}[1]{}\providecommand{\singleletter}[1]{#1}%
\begin{thebibliography}{49}%
\makeatletter
\providecommand \@ifxundefined [1]{%
 \@ifx{#1\undefined}
}%
\providecommand \@ifnum [1]{%
 \ifnum #1\expandafter \@firstoftwo
 \else \expandafter \@secondoftwo
 \fi
}%
\providecommand \@ifx [1]{%
 \ifx #1\expandafter \@firstoftwo
 \else \expandafter \@secondoftwo
 \fi
}%
\providecommand \natexlab [1]{#1}%
\providecommand \enquote  [1]{``#1''}%
\providecommand \bibnamefont  [1]{#1}%
\providecommand \bibfnamefont [1]{#1}%
\providecommand \citenamefont [1]{#1}%
\providecommand \href@noop [0]{\@secondoftwo}%
\providecommand \href [0]{\begingroup \@sanitize@url \@href}%
\providecommand \@href[1]{\@@startlink{#1}\@@href}%
\providecommand \@@href[1]{\endgroup#1\@@endlink}%
\providecommand \@sanitize@url [0]{\catcode `\\12\catcode `\$12\catcode
  `\&12\catcode `\#12\catcode `\^12\catcode `\_12\catcode `\%12\relax}%
\providecommand \@@startlink[1]{}%
\providecommand \@@endlink[0]{}%
\providecommand \url  [0]{\begingroup\@sanitize@url \@url }%
\providecommand \@url [1]{\endgroup\@href {#1}{\urlprefix }}%
\providecommand \urlprefix  [0]{URL }%
\providecommand \Eprint [0]{\href }%
\providecommand \doibase [0]{https://doi.org/}%
\providecommand \selectlanguage [0]{\@gobble}%
\providecommand \bibinfo  [0]{\@secondoftwo}%
\providecommand \bibfield  [0]{\@secondoftwo}%
\providecommand \translation [1]{[#1]}%
\providecommand \BibitemOpen [0]{}%
\providecommand \bibitemStop [0]{}%
\providecommand \bibitemNoStop [0]{.\EOS\space}%
\providecommand \EOS [0]{\spacefactor3000\relax}%
\providecommand \BibitemShut  [1]{\csname bibitem#1\endcsname}%
\let\auto@bib@innerbib\@empty
\bibitem [{\citenamefont {Busza}\ \emph {et~al.}(2018)\citenamefont {Busza},
  \citenamefont {Rajagopal},\ and\ \citenamefont {Van
  Der~Schee}}]{busza2018heavy}%
  \BibitemOpen
  \bibfield  {author} {\bibinfo {author} {\bibfnamefont {W.}~\bibnamefont
  {Busza}}, \bibinfo {author} {\bibfnamefont {K.}~\bibnamefont {Rajagopal}},\
  and\ \bibinfo {author} {\bibfnamefont {W.}~\bibnamefont {Van Der~Schee}},\
  }\bibfield  {title} {\bibinfo {title} {Heavy ion collisions: the big picture
  and the big questions},\ }\href@noop {} {\bibfield  {journal} {\bibinfo
  {journal} {Annual Review of Nuclear and Particle Science}\ }\textbf {\bibinfo
  {volume} {68}},\ \bibinfo {pages} {339} (\bibinfo {year} {2018})}\BibitemShut
  {NoStop}%
\bibitem [{\citenamefont {Philipsen}(2013)}]{philipsen2013qcd}%
  \BibitemOpen
  \bibfield  {author} {\bibinfo {author} {\bibfnamefont {O.}~\bibnamefont
  {Philipsen}},\ }\bibfield  {title} {\bibinfo {title} {The qcd equation of
  state from the lattice},\ }\href@noop {} {\bibfield  {journal} {\bibinfo
  {journal} {Progress in Particle and Nuclear Physics}\ }\textbf {\bibinfo
  {volume} {70}},\ \bibinfo {pages} {55} (\bibinfo {year} {2013})}\BibitemShut
  {NoStop}%
\bibitem [{\citenamefont {De~Forcrand}\ \emph {et~al.}(2001)\citenamefont
  {De~Forcrand}, \citenamefont {D'Elia},\ and\ \citenamefont {Pepe}}]{de2001t}%
  \BibitemOpen
  \bibfield  {author} {\bibinfo {author} {\bibfnamefont {P.}~\bibnamefont
  {De~Forcrand}}, \bibinfo {author} {\bibfnamefont {M.}~\bibnamefont
  {D'Elia}},\ and\ \bibinfo {author} {\bibfnamefont {M.}~\bibnamefont {Pepe}},\
  }\bibfield  {title} {\bibinfo {title} {'t hooft loop in su(2) yang-mills
  theory},\ }\href@noop {} {\bibfield  {journal} {\bibinfo  {journal} {Physical
  Review Letters}\ }\textbf {\bibinfo {volume} {86}},\ \bibinfo {pages} {1438}
  (\bibinfo {year} {2001})}\BibitemShut {NoStop}%
\bibitem [{\citenamefont {Giusti}\ and\ \citenamefont
  {Pepe}(2017)}]{giusti2017equation}%
  \BibitemOpen
  \bibfield  {author} {\bibinfo {author} {\bibfnamefont {L.}~\bibnamefont
  {Giusti}}\ and\ \bibinfo {author} {\bibfnamefont {M.}~\bibnamefont {Pepe}},\
  }\bibfield  {title} {\bibinfo {title} {Equation of state of the su (3)
  yang--mills theory: A precise determination from a moving frame},\
  }\href@noop {} {\bibfield  {journal} {\bibinfo  {journal} {Physics Letters
  B}\ }\textbf {\bibinfo {volume} {769}},\ \bibinfo {pages} {385} (\bibinfo
  {year} {2017})}\BibitemShut {NoStop}%
\bibitem [{\citenamefont {Caselle}\ \emph {et~al.}(2018)\citenamefont
  {Caselle}, \citenamefont {Nada},\ and\ \citenamefont
  {Panero}}]{caselle2018qcd}%
  \BibitemOpen
  \bibfield  {author} {\bibinfo {author} {\bibfnamefont {M.}~\bibnamefont
  {Caselle}}, \bibinfo {author} {\bibfnamefont {A.}~\bibnamefont {Nada}},\ and\
  \bibinfo {author} {\bibfnamefont {M.}~\bibnamefont {Panero}},\ }\bibfield
  {title} {\bibinfo {title} {Qcd thermodynamics from lattice calculations with
  nonequilibrium methods: The su(3) equation of state},\ }\href@noop {}
  {\bibfield  {journal} {\bibinfo  {journal} {Physical Review D}\ }\textbf
  {\bibinfo {volume} {98}},\ \bibinfo {pages} {054513} (\bibinfo {year}
  {2018})}\BibitemShut {NoStop}%
\bibitem [{\citenamefont
  {Jarzynski}(1997{\natexlab{a}})}]{jarzynski1997equilibrium}%
  \BibitemOpen
  \bibfield  {author} {\bibinfo {author} {\bibfnamefont {C.}~\bibnamefont
  {Jarzynski}},\ }\bibfield  {title} {\bibinfo {title} {Equilibrium free-energy
  differences from nonequilibrium measurements: A master-equation approach},\
  }\href@noop {} {\bibfield  {journal} {\bibinfo  {journal} {Physical Review
  E}\ }\textbf {\bibinfo {volume} {56}},\ \bibinfo {pages} {5018} (\bibinfo
  {year} {1997}{\natexlab{a}})}\BibitemShut {NoStop}%
\bibitem [{\citenamefont
  {Jarzynski}(1997{\natexlab{b}})}]{jarzynski1997nonequilibrium}%
  \BibitemOpen
  \bibfield  {author} {\bibinfo {author} {\bibfnamefont {C.}~\bibnamefont
  {Jarzynski}},\ }\bibfield  {title} {\bibinfo {title} {Nonequilibrium equality
  for free energy differences},\ }\href@noop {} {\bibfield  {journal} {\bibinfo
   {journal} {Physical Review Letters}\ }\textbf {\bibinfo {volume} {78}},\
  \bibinfo {pages} {2690} (\bibinfo {year} {1997}{\natexlab{b}})}\BibitemShut
  {NoStop}%
\bibitem [{\citenamefont {Goodfellow}\ \emph {et~al.}(2016)\citenamefont
  {Goodfellow}, \citenamefont {Bengio},\ and\ \citenamefont
  {Courville}}]{goodfellow2016deep}%
  \BibitemOpen
  \bibfield  {author} {\bibinfo {author} {\bibfnamefont {I.}~\bibnamefont
  {Goodfellow}}, \bibinfo {author} {\bibfnamefont {Y.}~\bibnamefont {Bengio}},\
  and\ \bibinfo {author} {\bibfnamefont {A.}~\bibnamefont {Courville}},\
  }\href@noop {} {\emph {\bibinfo {title} {Deep learning}}}\ (\bibinfo
  {publisher} {MIT press},\ \bibinfo {year} {2016})\BibitemShut {NoStop}%
\bibitem [{\citenamefont {Shanahan}\ \emph {et~al.}(2018)\citenamefont
  {Shanahan}, \citenamefont {Trewartha},\ and\ \citenamefont
  {Detmold}}]{shanahan2018machine}%
  \BibitemOpen
  \bibfield  {author} {\bibinfo {author} {\bibfnamefont {P.~E.}\ \bibnamefont
  {Shanahan}}, \bibinfo {author} {\bibfnamefont {D.}~\bibnamefont
  {Trewartha}},\ and\ \bibinfo {author} {\bibfnamefont {W.}~\bibnamefont
  {Detmold}},\ }\bibfield  {title} {\bibinfo {title} {Machine learning action
  parameters in lattice quantum chromodynamics},\ }\href@noop {} {\bibfield
  {journal} {\bibinfo  {journal} {Physical Review D}\ }\textbf {\bibinfo
  {volume} {97}},\ \bibinfo {pages} {094506} (\bibinfo {year}
  {2018})}\BibitemShut {NoStop}%
\bibitem [{\citenamefont {Bachtis}\ \emph {et~al.}(2020)\citenamefont
  {Bachtis}, \citenamefont {Aarts},\ and\ \citenamefont
  {Lucini}}]{bachtis2020mapping}%
  \BibitemOpen
  \bibfield  {author} {\bibinfo {author} {\bibfnamefont {D.}~\bibnamefont
  {Bachtis}}, \bibinfo {author} {\bibfnamefont {G.}~\bibnamefont {Aarts}},\
  and\ \bibinfo {author} {\bibfnamefont {B.}~\bibnamefont {Lucini}},\
  }\bibfield  {title} {\bibinfo {title} {A mapping of distinct phase
  transitions to a neural network},\ }\href@noop {} {\bibfield  {journal}
  {\bibinfo  {journal} {arXiv preprint arXiv:2007.00355}\ } (\bibinfo {year}
  {2020})}\BibitemShut {NoStop}%
\bibitem [{\citenamefont {Lee}(2020)}]{hol1}%
  \BibitemOpen
  \bibfield  {author} {\bibinfo {author} {\bibfnamefont {J.-W.}\ \bibnamefont
  {Lee}},\ }\bibfield  {title} {\bibinfo {title} {Quantum fields as deep
  learning},\ }\href@noop {} {\bibfield  {journal} {\bibinfo  {journal}
  {Journal of the Korean Physical Society}\ }\textbf {\bibinfo {volume} {76}},\
  \bibinfo {pages} {684} (\bibinfo {year} {2020})}\BibitemShut {NoStop}%
\bibitem [{\citenamefont {Hashimoto}\ \emph
  {et~al.}(2018{\natexlab{a}})\citenamefont {Hashimoto}, \citenamefont
  {Sugishita}, \citenamefont {Tanaka},\ and\ \citenamefont {Tomiya}}]{hol2}%
  \BibitemOpen
  \bibfield  {author} {\bibinfo {author} {\bibfnamefont {K.}~\bibnamefont
  {Hashimoto}}, \bibinfo {author} {\bibfnamefont {S.}~\bibnamefont
  {Sugishita}}, \bibinfo {author} {\bibfnamefont {A.}~\bibnamefont {Tanaka}},\
  and\ \bibinfo {author} {\bibfnamefont {A.}~\bibnamefont {Tomiya}},\
  }\bibfield  {title} {\bibinfo {title} {Deep learning and the ads/cft
  correspondence},\ }\href@noop {} {\bibfield  {journal} {\bibinfo  {journal}
  {Physical Review D}\ }\textbf {\bibinfo {volume} {98}},\ \bibinfo {pages}
  {046019} (\bibinfo {year} {2018}{\natexlab{a}})}\BibitemShut {NoStop}%
\bibitem [{\citenamefont {Hashimoto}\ \emph
  {et~al.}(2018{\natexlab{b}})\citenamefont {Hashimoto}, \citenamefont
  {Sugishita}, \citenamefont {Tanaka},\ and\ \citenamefont {Tomiya}}]{hol3}%
  \BibitemOpen
  \bibfield  {author} {\bibinfo {author} {\bibfnamefont {K.}~\bibnamefont
  {Hashimoto}}, \bibinfo {author} {\bibfnamefont {S.}~\bibnamefont
  {Sugishita}}, \bibinfo {author} {\bibfnamefont {A.}~\bibnamefont {Tanaka}},\
  and\ \bibinfo {author} {\bibfnamefont {A.}~\bibnamefont {Tomiya}},\
  }\bibfield  {title} {\bibinfo {title} {Deep learning and holographic qcd},\
  }\href@noop {} {\bibfield  {journal} {\bibinfo  {journal} {Physical Review
  D}\ }\textbf {\bibinfo {volume} {98}},\ \bibinfo {pages} {106014} (\bibinfo
  {year} {2018}{\natexlab{b}})}\BibitemShut {NoStop}%
\bibitem [{\citenamefont {Hashimoto}(2019)}]{hol4}%
  \BibitemOpen
  \bibfield  {author} {\bibinfo {author} {\bibfnamefont {K.}~\bibnamefont
  {Hashimoto}},\ }\bibfield  {title} {\bibinfo {title} {Ads/cft correspondence
  as a deep boltzmann machine},\ }\href@noop {} {\bibfield  {journal} {\bibinfo
   {journal} {Physical Review D}\ }\textbf {\bibinfo {volume} {99}},\ \bibinfo
  {pages} {106017} (\bibinfo {year} {2019})}\BibitemShut {NoStop}%
\bibitem [{\citenamefont {Akutagawa}\ \emph {et~al.}(2020)\citenamefont
  {Akutagawa}, \citenamefont {Hashimoto},\ and\ \citenamefont
  {Sumimoto}}]{hol5}%
  \BibitemOpen
  \bibfield  {author} {\bibinfo {author} {\bibfnamefont {T.}~\bibnamefont
  {Akutagawa}}, \bibinfo {author} {\bibfnamefont {K.}~\bibnamefont
  {Hashimoto}},\ and\ \bibinfo {author} {\bibfnamefont {T.}~\bibnamefont
  {Sumimoto}},\ }\bibfield  {title} {\bibinfo {title} {Deep learning and
  ads/qcd},\ }\href@noop {} {\bibfield  {journal} {\bibinfo  {journal} {arXiv
  preprint arXiv:2005.02636}\ } (\bibinfo {year} {2020})}\BibitemShut {NoStop}%
\bibitem [{\citenamefont {Albergo}\ \emph {et~al.}(2019)\citenamefont
  {Albergo}, \citenamefont {Kanwar},\ and\ \citenamefont
  {Shanahan}}]{flowsforlattice}%
  \BibitemOpen
  \bibfield  {author} {\bibinfo {author} {\bibfnamefont {M.~S.}\ \bibnamefont
  {Albergo}}, \bibinfo {author} {\bibfnamefont {G.}~\bibnamefont {Kanwar}},\
  and\ \bibinfo {author} {\bibfnamefont {P.~E.}\ \bibnamefont {Shanahan}},\
  }\bibfield  {title} {\bibinfo {title} {Flow-based generative models for
  markov chain monte carlo in lattice field theory},\ }\href
  {https://doi.org/10.1103/PhysRevD.100.034515} {\bibfield  {journal} {\bibinfo
   {journal} {Phys. Rev. D}\ }\textbf {\bibinfo {volume} {100}},\ \bibinfo
  {pages} {034515} (\bibinfo {year} {2019})}\BibitemShut {NoStop}%
\bibitem [{\citenamefont {Kanwar}\ \emph {et~al.}(2020)\citenamefont {Kanwar},
  \citenamefont {Albergo}, \citenamefont {Boyda}, \citenamefont {Cranmer},
  \citenamefont {Hackett}, \citenamefont {Racani\`ere}, \citenamefont
  {Rezende},\ and\ \citenamefont {Shanahan}}]{flowsforlattice2}%
  \BibitemOpen
  \bibfield  {author} {\bibinfo {author} {\bibfnamefont {G.}~\bibnamefont
  {Kanwar}}, \bibinfo {author} {\bibfnamefont {M.~S.}\ \bibnamefont {Albergo}},
  \bibinfo {author} {\bibfnamefont {D.}~\bibnamefont {Boyda}}, \bibinfo
  {author} {\bibfnamefont {K.}~\bibnamefont {Cranmer}}, \bibinfo {author}
  {\bibfnamefont {D.~C.}\ \bibnamefont {Hackett}}, \bibinfo {author}
  {\bibfnamefont {S.}~\bibnamefont {Racani\`ere}}, \bibinfo {author}
  {\bibfnamefont {D.~J.}\ \bibnamefont {Rezende}},\ and\ \bibinfo {author}
  {\bibfnamefont {P.~E.}\ \bibnamefont {Shanahan}},\ }\bibfield  {title}
  {\bibinfo {title} {Equivariant flow-based sampling for lattice gauge
  theory},\ }\href@noop {} {\bibfield  {journal} {\bibinfo  {journal} {Phys.
  Rev. Lett.}\ }\textbf {\bibinfo {volume} {125}},\ \bibinfo {pages} {121601}
  (\bibinfo {year} {2020})}\BibitemShut {NoStop}%
\bibitem [{\citenamefont {Boyda}\ \emph {et~al.}(2020)\citenamefont {Boyda},
  \citenamefont {Kanwar}, \citenamefont {Racani{\`e}re}, \citenamefont
  {Rezende}, \citenamefont {Albergo}, \citenamefont {Cranmer}, \citenamefont
  {Hackett},\ and\ \citenamefont {Shanahan}}]{flowsforlattice3}%
  \BibitemOpen
  \bibfield  {author} {\bibinfo {author} {\bibfnamefont {D.}~\bibnamefont
  {Boyda}}, \bibinfo {author} {\bibfnamefont {G.}~\bibnamefont {Kanwar}},
  \bibinfo {author} {\bibfnamefont {S.}~\bibnamefont {Racani{\`e}re}}, \bibinfo
  {author} {\bibfnamefont {D.~J.}\ \bibnamefont {Rezende}}, \bibinfo {author}
  {\bibfnamefont {M.~S.}\ \bibnamefont {Albergo}}, \bibinfo {author}
  {\bibfnamefont {K.}~\bibnamefont {Cranmer}}, \bibinfo {author} {\bibfnamefont
  {D.~C.}\ \bibnamefont {Hackett}},\ and\ \bibinfo {author} {\bibfnamefont
  {P.~E.}\ \bibnamefont {Shanahan}},\ }\bibfield  {title} {\bibinfo {title}
  {Sampling using $ su (n) $ gauge equivariant flows},\ }\href@noop {}
  {\bibfield  {journal} {\bibinfo  {journal} {arXiv preprint arXiv:2008.05456}\
  } (\bibinfo {year} {2020})}\BibitemShut {NoStop}%
\bibitem [{\citenamefont {Pawlowski}\ and\ \citenamefont
  {Urban}(2020)}]{urban2018reducing}%
  \BibitemOpen
  \bibfield  {author} {\bibinfo {author} {\bibfnamefont {J.~M.}\ \bibnamefont
  {Pawlowski}}\ and\ \bibinfo {author} {\bibfnamefont {J.~M.}\ \bibnamefont
  {Urban}},\ }\bibfield  {title} {\bibinfo {title} {Reducing autocorrelation
  times in lattice simulations with generative adversarial networks},\
  }\href@noop {} {\bibfield  {journal} {\bibinfo  {journal} {Machine Learning:
  Science and Technology}\ }\textbf {\bibinfo {volume} {1}},\ \bibinfo {pages}
  {045011} (\bibinfo {year} {2020})}\BibitemShut {NoStop}%
\bibitem [{\citenamefont {Tanaka}\ and\ \citenamefont
  {Tomiya}(2017)}]{tanaka2017towards}%
  \BibitemOpen
  \bibfield  {author} {\bibinfo {author} {\bibfnamefont {A.}~\bibnamefont
  {Tanaka}}\ and\ \bibinfo {author} {\bibfnamefont {A.}~\bibnamefont
  {Tomiya}},\ }\bibfield  {title} {\bibinfo {title} {Towards reduction of
  autocorrelation in hmc by machine learning},\ }\href@noop {} {\bibfield
  {journal} {\bibinfo  {journal} {arXiv preprint arXiv:1712.03893}\ } (\bibinfo
  {year} {2017})}\BibitemShut {NoStop}%
\bibitem [{\citenamefont {Wirnsberger}\ \emph {et~al.}(2020)\citenamefont
  {Wirnsberger}, \citenamefont {Ballard}, \citenamefont {Papamakarios},
  \citenamefont {Abercrombie}, \citenamefont {Racani{\`e}re}, \citenamefont
  {Pritzel}, \citenamefont {Rezende},\ and\ \citenamefont
  {Blundell}}]{wirnsberger2020targeted}%
  \BibitemOpen
  \bibfield  {author} {\bibinfo {author} {\bibfnamefont {P.}~\bibnamefont
  {Wirnsberger}}, \bibinfo {author} {\bibfnamefont {A.~J.}\ \bibnamefont
  {Ballard}}, \bibinfo {author} {\bibfnamefont {G.}~\bibnamefont
  {Papamakarios}}, \bibinfo {author} {\bibfnamefont {S.}~\bibnamefont
  {Abercrombie}}, \bibinfo {author} {\bibfnamefont {S.}~\bibnamefont
  {Racani{\`e}re}}, \bibinfo {author} {\bibfnamefont {A.}~\bibnamefont
  {Pritzel}}, \bibinfo {author} {\bibfnamefont {D.~J.}\ \bibnamefont
  {Rezende}},\ and\ \bibinfo {author} {\bibfnamefont {C.}~\bibnamefont
  {Blundell}},\ }\bibfield  {title} {\bibinfo {title} {Targeted free energy
  estimation via learned mappings},\ }\href@noop {} {\bibfield  {journal}
  {\bibinfo  {journal} {arXiv preprint arXiv:2002.04913}\ } (\bibinfo {year}
  {2020})}\BibitemShut {NoStop}%
\bibitem [{\citenamefont {Zwanzig}(1955)}]{zwanzig1955high}%
  \BibitemOpen
  \bibfield  {author} {\bibinfo {author} {\bibfnamefont {R.~W.}\ \bibnamefont
  {Zwanzig}},\ }\bibfield  {title} {\bibinfo {title} {High-temperature equation
  of state by a perturbation method. ii. polar gases},\ }\href@noop {}
  {\bibfield  {journal} {\bibinfo  {journal} {The Journal of Chemical Physics}\
  }\textbf {\bibinfo {volume} {23}},\ \bibinfo {pages} {1915} (\bibinfo {year}
  {1955})}\BibitemShut {NoStop}%
\bibitem [{\citenamefont {Akiyama}\ \emph {et~al.}(2020)\citenamefont
  {Akiyama}, \citenamefont {Kadoh}, \citenamefont {Kuramashi}, \citenamefont
  {Yamashita},\ and\ \citenamefont {Yoshimura}}]{akiyama2020tensor}%
  \BibitemOpen
  \bibfield  {author} {\bibinfo {author} {\bibfnamefont {S.}~\bibnamefont
  {Akiyama}}, \bibinfo {author} {\bibfnamefont {D.}~\bibnamefont {Kadoh}},
  \bibinfo {author} {\bibfnamefont {Y.}~\bibnamefont {Kuramashi}}, \bibinfo
  {author} {\bibfnamefont {T.}~\bibnamefont {Yamashita}},\ and\ \bibinfo
  {author} {\bibfnamefont {Y.}~\bibnamefont {Yoshimura}},\ }\bibfield  {title}
  {\bibinfo {title} {Tensor renormalization group approach to four-dimensional
  complex $\phi^4$ theory at finite density},\ }\href@noop {} {\bibfield
  {journal} {\bibinfo  {journal} {arXiv preprint arXiv:2005.04645}\ } (\bibinfo
  {year} {2020})}\BibitemShut {NoStop}%
\bibitem [{\citenamefont {Asakawa}\ \emph {et~al.}(2014)\citenamefont
  {Asakawa}, \citenamefont {Hatsuda}, \citenamefont {Itou}, \citenamefont
  {Kitazawa}, \citenamefont {Suzuki}, \citenamefont {Collaboration} \emph
  {et~al.}}]{asakawa2014thermodynamics}%
  \BibitemOpen
  \bibfield  {author} {\bibinfo {author} {\bibfnamefont {M.}~\bibnamefont
  {Asakawa}}, \bibinfo {author} {\bibfnamefont {T.}~\bibnamefont {Hatsuda}},
  \bibinfo {author} {\bibfnamefont {E.}~\bibnamefont {Itou}}, \bibinfo {author}
  {\bibfnamefont {M.}~\bibnamefont {Kitazawa}}, \bibinfo {author}
  {\bibfnamefont {H.}~\bibnamefont {Suzuki}}, \bibinfo {author} {\bibfnamefont
  {F.}~\bibnamefont {Collaboration}}, \emph {et~al.},\ }\bibfield  {title}
  {\bibinfo {title} {Thermodynamics of s u (3) gauge theory from gradient flow
  on the lattice},\ }\href@noop {} {\bibfield  {journal} {\bibinfo  {journal}
  {Physical Review D}\ }\textbf {\bibinfo {volume} {90}},\ \bibinfo {pages}
  {011501} (\bibinfo {year} {2014})}\BibitemShut {NoStop}%
\bibitem [{\citenamefont {Giusti}\ and\ \citenamefont
  {Pepe}(2014)}]{giusti2014equation}%
  \BibitemOpen
  \bibfield  {author} {\bibinfo {author} {\bibfnamefont {L.}~\bibnamefont
  {Giusti}}\ and\ \bibinfo {author} {\bibfnamefont {M.}~\bibnamefont {Pepe}},\
  }\bibfield  {title} {\bibinfo {title} {Equation of state of a relativistic
  theory from a moving frame},\ }\href@noop {} {\bibfield  {journal} {\bibinfo
  {journal} {Physical review letters}\ }\textbf {\bibinfo {volume} {113}},\
  \bibinfo {pages} {031601} (\bibinfo {year} {2014})}\BibitemShut {NoStop}%
\bibitem [{\citenamefont {Papamakarios}\ \emph {et~al.}(2019)\citenamefont
  {Papamakarios}, \citenamefont {Nalisnick}, \citenamefont {Rezende},
  \citenamefont {Mohamed},\ and\ \citenamefont {Lakshminarayanan}}]{nfreview}%
  \BibitemOpen
  \bibfield  {author} {\bibinfo {author} {\bibfnamefont {G.}~\bibnamefont
  {Papamakarios}}, \bibinfo {author} {\bibfnamefont {E.}~\bibnamefont
  {Nalisnick}}, \bibinfo {author} {\bibfnamefont {D.~J.}\ \bibnamefont
  {Rezende}}, \bibinfo {author} {\bibfnamefont {S.}~\bibnamefont {Mohamed}},\
  and\ \bibinfo {author} {\bibfnamefont {B.}~\bibnamefont {Lakshminarayanan}},\
  }\bibfield  {title} {\bibinfo {title} {Normalizing flows for probabilistic
  modeling and inference},\ }\href@noop {} {\bibfield  {journal} {\bibinfo
  {journal} {arXiv preprint arXiv:1912.02762}\ } (\bibinfo {year}
  {2019})}\BibitemShut {NoStop}%
\bibitem [{\citenamefont {Dinh}\ \emph {et~al.}(2014)\citenamefont {Dinh},
  \citenamefont {Krueger},\ and\ \citenamefont {Bengio}}]{nice}%
  \BibitemOpen
  \bibfield  {author} {\bibinfo {author} {\bibfnamefont {L.}~\bibnamefont
  {Dinh}}, \bibinfo {author} {\bibfnamefont {D.}~\bibnamefont {Krueger}},\ and\
  \bibinfo {author} {\bibfnamefont {Y.}~\bibnamefont {Bengio}},\ }\bibfield
  {title} {\bibinfo {title} {Nice: Non-linear independent components
  estimation},\ }\href@noop {} {\bibfield  {journal} {\bibinfo  {journal}
  {arXiv preprint arXiv:1410.8516}\ } (\bibinfo {year} {2014})}\BibitemShut
  {NoStop}%
\bibitem [{\citenamefont {MacKay}\ and\ \citenamefont
  {Mac~Kay}(2003)}]{mackay2003information}%
  \BibitemOpen
  \bibfield  {author} {\bibinfo {author} {\bibfnamefont {D.~J.}\ \bibnamefont
  {MacKay}}\ and\ \bibinfo {author} {\bibfnamefont {D.~J.}\ \bibnamefont
  {Mac~Kay}},\ }\href@noop {} {\emph {\bibinfo {title} {Information theory,
  inference and learning algorithms}}}\ (\bibinfo  {publisher} {Cambridge
  university press},\ \bibinfo {year} {2003})\BibitemShut {NoStop}%
\bibitem [{Note1()}]{Note1}%
  \BibitemOpen
  \bibinfo {note} {More precisely, the distributions $p$ and $q$ have to be
  identically only almost everywhere, i.e. up to a set of measure
  zero.}\BibitemShut {Stop}%
\bibitem [{\citenamefont {Kingma}\ and\ \citenamefont
  {Welling}(2013)}]{kingma2013auto}%
  \BibitemOpen
  \bibfield  {author} {\bibinfo {author} {\bibfnamefont {D.~P.}\ \bibnamefont
  {Kingma}}\ and\ \bibinfo {author} {\bibfnamefont {M.}~\bibnamefont
  {Welling}},\ }\bibfield  {title} {\bibinfo {title} {Auto-encoding variational
  bayes},\ }\href@noop {} {\bibfield  {journal} {\bibinfo  {journal} {arXiv
  preprint arXiv:1312.6114}\ } (\bibinfo {year} {2013})}\BibitemShut {NoStop}%
\bibitem [{\citenamefont {Rezende}\ \emph {et~al.}(2014)\citenamefont
  {Rezende}, \citenamefont {Mohamed},\ and\ \citenamefont
  {Wierstra}}]{rezende2014stochastic}%
  \BibitemOpen
  \bibfield  {author} {\bibinfo {author} {\bibfnamefont {D.~J.}\ \bibnamefont
  {Rezende}}, \bibinfo {author} {\bibfnamefont {S.}~\bibnamefont {Mohamed}},\
  and\ \bibinfo {author} {\bibfnamefont {D.}~\bibnamefont {Wierstra}},\
  }\bibfield  {title} {\bibinfo {title} {Stochastic backpropagation and
  approximate inference in deep generative models},\ }\href@noop {} {\bibfield
  {journal} {\bibinfo  {journal} {arXiv preprint arXiv:1401.4082}\ } (\bibinfo
  {year} {2014})}\BibitemShut {NoStop}%
\bibitem [{\citenamefont {Rezende}\ and\ \citenamefont
  {Mohamed}(2015)}]{rezende2015variational}%
  \BibitemOpen
  \bibfield  {author} {\bibinfo {author} {\bibfnamefont {D.~J.}\ \bibnamefont
  {Rezende}}\ and\ \bibinfo {author} {\bibfnamefont {S.}~\bibnamefont
  {Mohamed}},\ }\bibfield  {title} {\bibinfo {title} {Variational inference
  with normalizing flows},\ }\href@noop {} {\bibfield  {journal} {\bibinfo
  {journal} {arXiv preprint arXiv:1505.05770}\ } (\bibinfo {year}
  {2015})}\BibitemShut {NoStop}%
\bibitem [{\citenamefont {M{\"u}ller}\ \emph {et~al.}(2019)\citenamefont
  {M{\"u}ller}, \citenamefont {Mcwilliams}, \citenamefont {Rousselle},
  \citenamefont {Gross},\ and\ \citenamefont {Nov{\'a}k}}]{muller2019neural}%
  \BibitemOpen
  \bibfield  {author} {\bibinfo {author} {\bibfnamefont {T.}~\bibnamefont
  {M{\"u}ller}}, \bibinfo {author} {\bibfnamefont {B.}~\bibnamefont
  {Mcwilliams}}, \bibinfo {author} {\bibfnamefont {F.}~\bibnamefont
  {Rousselle}}, \bibinfo {author} {\bibfnamefont {M.}~\bibnamefont {Gross}},\
  and\ \bibinfo {author} {\bibfnamefont {J.}~\bibnamefont {Nov{\'a}k}},\
  }\bibfield  {title} {\bibinfo {title} {Neural importance sampling},\
  }\href@noop {} {\bibfield  {journal} {\bibinfo  {journal} {ACM Transactions
  on Graphics (TOG)}\ }\textbf {\bibinfo {volume} {38}},\ \bibinfo {pages} {1}
  (\bibinfo {year} {2019})}\BibitemShut {NoStop}%
\bibitem [{\citenamefont {No{\'e}}\ \emph {et~al.}(2019)\citenamefont
  {No{\'e}}, \citenamefont {Olsson}, \citenamefont {K{\"o}hler},\ and\
  \citenamefont {Wu}}]{noe2019boltzmann}%
  \BibitemOpen
  \bibfield  {author} {\bibinfo {author} {\bibfnamefont {F.}~\bibnamefont
  {No{\'e}}}, \bibinfo {author} {\bibfnamefont {S.}~\bibnamefont {Olsson}},
  \bibinfo {author} {\bibfnamefont {J.}~\bibnamefont {K{\"o}hler}},\ and\
  \bibinfo {author} {\bibfnamefont {H.}~\bibnamefont {Wu}},\ }\bibfield
  {title} {\bibinfo {title} {Boltzmann generators: Sampling equilibrium states
  of many-body systems with deep learning},\ }\href@noop {} {\bibfield
  {journal} {\bibinfo  {journal} {Science}\ }\textbf {\bibinfo {volume} {365}}
  (\bibinfo {year} {2019})}\BibitemShut {NoStop}%
\bibitem [{\citenamefont {Wu}\ \emph {et~al.}(2019)\citenamefont {Wu},
  \citenamefont {Wang},\ and\ \citenamefont {Zhang}}]{wu2019solving}%
  \BibitemOpen
  \bibfield  {author} {\bibinfo {author} {\bibfnamefont {D.}~\bibnamefont
  {Wu}}, \bibinfo {author} {\bibfnamefont {L.}~\bibnamefont {Wang}},\ and\
  \bibinfo {author} {\bibfnamefont {P.}~\bibnamefont {Zhang}},\ }\bibfield
  {title} {\bibinfo {title} {Solving statistical mechanics using variational
  autoregressive networks},\ }\href@noop {} {\bibfield  {journal} {\bibinfo
  {journal} {Physical Review Letters}\ }\textbf {\bibinfo {volume} {122}},\
  \bibinfo {pages} {080602} (\bibinfo {year} {2019})}\BibitemShut {NoStop}%
\bibitem [{\citenamefont {Nicoli}\ \emph {et~al.}(2020)\citenamefont {Nicoli},
  \citenamefont {Nakajima}, \citenamefont {Strodthoff}, \citenamefont {Samek},
  \citenamefont {M{\"u}ller},\ and\ \citenamefont
  {Kessel}}]{nicoli2020asymptotically}%
  \BibitemOpen
  \bibfield  {author} {\bibinfo {author} {\bibfnamefont {K.~A.}\ \bibnamefont
  {Nicoli}}, \bibinfo {author} {\bibfnamefont {S.}~\bibnamefont {Nakajima}},
  \bibinfo {author} {\bibfnamefont {N.}~\bibnamefont {Strodthoff}}, \bibinfo
  {author} {\bibfnamefont {W.}~\bibnamefont {Samek}}, \bibinfo {author}
  {\bibfnamefont {K.-R.}\ \bibnamefont {M{\"u}ller}},\ and\ \bibinfo {author}
  {\bibfnamefont {P.}~\bibnamefont {Kessel}},\ }\bibfield  {title} {\bibinfo
  {title} {Asymptotically unbiased estimation of physical observables with
  neural samplers},\ }\href@noop {} {\bibfield  {journal} {\bibinfo  {journal}
  {Physical Review E}\ }\textbf {\bibinfo {volume} {101}},\ \bibinfo {pages}
  {023304} (\bibinfo {year} {2020})}\BibitemShut {NoStop}%
\bibitem [{\citenamefont {Nicoli}\ \emph {et~al.}(2019)\citenamefont {Nicoli},
  \citenamefont {Kessel}, \citenamefont {Strodthoff}, \citenamefont {Samek},
  \citenamefont {M{\"u}ller},\ and\ \citenamefont
  {Nakajima}}]{nicoli2019comment}%
  \BibitemOpen
  \bibfield  {author} {\bibinfo {author} {\bibfnamefont {K.}~\bibnamefont
  {Nicoli}}, \bibinfo {author} {\bibfnamefont {P.}~\bibnamefont {Kessel}},
  \bibinfo {author} {\bibfnamefont {N.}~\bibnamefont {Strodthoff}}, \bibinfo
  {author} {\bibfnamefont {W.}~\bibnamefont {Samek}}, \bibinfo {author}
  {\bibfnamefont {K.-R.}\ \bibnamefont {M{\"u}ller}},\ and\ \bibinfo {author}
  {\bibfnamefont {S.}~\bibnamefont {Nakajima}},\ }\bibfield  {title} {\bibinfo
  {title} {Comment on "solving statistical mechanics using vans": Introducing
  savant-vans enhanced by importance and mcmc sampling},\ }\href@noop {}
  {\bibfield  {journal} {\bibinfo  {journal} {arXiv preprint arXiv:1903.11048}\
  } (\bibinfo {year} {2019})}\BibitemShut {NoStop}%
\bibitem [{\citenamefont {Adler}(1981)}]{adler1981over}%
  \BibitemOpen
  \bibfield  {author} {\bibinfo {author} {\bibfnamefont {S.~L.}\ \bibnamefont
  {Adler}},\ }\bibfield  {title} {\bibinfo {title} {Over-relaxation method for
  the monte carlo evaluation of the partition function for multiquadratic
  actions},\ }\href@noop {} {\bibfield  {journal} {\bibinfo  {journal}
  {Physical Review D}\ }\textbf {\bibinfo {volume} {23}},\ \bibinfo {pages}
  {2901} (\bibinfo {year} {1981})}\BibitemShut {NoStop}%
\bibitem [{\citenamefont {Whitmer}(1984)}]{whitmer1984over}%
  \BibitemOpen
  \bibfield  {author} {\bibinfo {author} {\bibfnamefont {C.}~\bibnamefont
  {Whitmer}},\ }\bibfield  {title} {\bibinfo {title} {Over-relaxation methods
  for monte carlo simulations of quadratic and multiquadratic actions},\
  }\href@noop {} {\bibfield  {journal} {\bibinfo  {journal} {Physical Review
  D}\ }\textbf {\bibinfo {volume} {29}},\ \bibinfo {pages} {306} (\bibinfo
  {year} {1984})}\BibitemShut {NoStop}%
\bibitem [{\citenamefont {Callaway}\ and\ \citenamefont
  {Rahman}(1983)}]{callaway1983lattice}%
  \BibitemOpen
  \bibfield  {author} {\bibinfo {author} {\bibfnamefont {D.~J.}\ \bibnamefont
  {Callaway}}\ and\ \bibinfo {author} {\bibfnamefont {A.}~\bibnamefont
  {Rahman}},\ }\bibfield  {title} {\bibinfo {title} {Lattice gauge theory in
  the microcanonical ensemble},\ }\href@noop {} {\bibfield  {journal} {\bibinfo
   {journal} {Physical Review D}\ }\textbf {\bibinfo {volume} {28}},\ \bibinfo
  {pages} {1506} (\bibinfo {year} {1983})}\BibitemShut {NoStop}%
\bibitem [{\citenamefont {Fodor}\ and\ \citenamefont
  {Jansen}(1994)}]{fodor1994overrelaxation}%
  \BibitemOpen
  \bibfield  {author} {\bibinfo {author} {\bibfnamefont {Z.}~\bibnamefont
  {Fodor}}\ and\ \bibinfo {author} {\bibfnamefont {K.}~\bibnamefont {Jansen}},\
  }\bibfield  {title} {\bibinfo {title} {Overrelaxation algorithm for coupled
  gauge-higgs systems},\ }\href@noop {} {\bibfield  {journal} {\bibinfo
  {journal} {Physics Letters B}\ }\textbf {\bibinfo {volume} {331}},\ \bibinfo
  {pages} {119} (\bibinfo {year} {1994})}\BibitemShut {NoStop}%
\bibitem [{\citenamefont {Wolff}\ \emph {et~al.}(2004)\citenamefont {Wolff},
  \citenamefont {Collaboration} \emph {et~al.}}]{wolff2004monte}%
  \BibitemOpen
  \bibfield  {author} {\bibinfo {author} {\bibfnamefont {U.}~\bibnamefont
  {Wolff}}, \bibinfo {author} {\bibfnamefont {A.}~\bibnamefont
  {Collaboration}}, \emph {et~al.},\ }\bibfield  {title} {\bibinfo {title}
  {Monte carlo errors with less errors},\ }\href@noop {} {\bibfield  {journal}
  {\bibinfo  {journal} {Computer Physics Communications}\ }\textbf {\bibinfo
  {volume} {156}},\ \bibinfo {pages} {143} (\bibinfo {year}
  {2004})}\BibitemShut {NoStop}%
\bibitem [{\citenamefont {Burda}\ \emph {et~al.}(2015)\citenamefont {Burda},
  \citenamefont {Grosse},\ and\ \citenamefont
  {Salakhutdinov}}]{burda2015importance}%
  \BibitemOpen
  \bibfield  {author} {\bibinfo {author} {\bibfnamefont {Y.}~\bibnamefont
  {Burda}}, \bibinfo {author} {\bibfnamefont {R.}~\bibnamefont {Grosse}},\ and\
  \bibinfo {author} {\bibfnamefont {R.}~\bibnamefont {Salakhutdinov}},\
  }\bibfield  {title} {\bibinfo {title} {Importance weighted autoencoders},\
  }\href@noop {} {\bibfield  {journal} {\bibinfo  {journal} {arXiv preprint
  arXiv:1509.00519}\ } (\bibinfo {year} {2015})}\BibitemShut {NoStop}%
\bibitem [{\citenamefont {Nowozin}(2018)}]{nowozin2018debiasing}%
  \BibitemOpen
  \bibfield  {author} {\bibinfo {author} {\bibfnamefont {S.}~\bibnamefont
  {Nowozin}},\ }\bibfield  {title} {\bibinfo {title} {Debiasing evidence
  approximations: On importance-weighted autoencoders and jackknife variational
  inference},\ }\href@noop {} {\bibfield  {journal} {\bibinfo  {journal} {ICLR
  2018}\ } (\bibinfo {year} {2018})}\BibitemShut {NoStop}%
\bibitem [{\citenamefont {Teh}\ \emph {et~al.}(2007)\citenamefont {Teh},
  \citenamefont {Newman},\ and\ \citenamefont {Welling}}]{teh2007collapsed}%
  \BibitemOpen
  \bibfield  {author} {\bibinfo {author} {\bibfnamefont {Y.~W.}\ \bibnamefont
  {Teh}}, \bibinfo {author} {\bibfnamefont {D.}~\bibnamefont {Newman}},\ and\
  \bibinfo {author} {\bibfnamefont {M.}~\bibnamefont {Welling}},\ }\bibfield
  {title} {\bibinfo {title} {A collapsed variational bayesian inference
  algorithm for latent dirichlet allocation},\ }in\ \href@noop {} {\emph
  {\bibinfo {booktitle} {Advances in neural information processing systems}}}\
  (\bibinfo {year} {2007})\ pp.\ \bibinfo {pages} {1353--1360}\BibitemShut
  {NoStop}%
\bibitem [{\citenamefont {Gattringer}\ and\ \citenamefont
  {Lang}(2009)}]{gattringer2009quantum}%
  \BibitemOpen
  \bibfield  {author} {\bibinfo {author} {\bibfnamefont {C.}~\bibnamefont
  {Gattringer}}\ and\ \bibinfo {author} {\bibfnamefont {C.}~\bibnamefont
  {Lang}},\ }\href@noop {} {\emph {\bibinfo {title} {Quantum chromodynamics on
  the lattice: an introductory presentation}}},\ Vol.\ \bibinfo {volume} {788}\
  (\bibinfo  {publisher} {Springer Science \& Business Media},\ \bibinfo {year}
  {2009})\BibitemShut {NoStop}%
\bibitem [{\citenamefont {Bishop}(2006)}]{bishop2006pattern}%
  \BibitemOpen
  \bibfield  {author} {\bibinfo {author} {\bibfnamefont {C.~M.}\ \bibnamefont
  {Bishop}},\ }\href@noop {} {\emph {\bibinfo {title} {Pattern recognition and
  machine learning}}}\ (\bibinfo  {publisher} {springer},\ \bibinfo {year}
  {2006})\ pp.\ \bibinfo {pages} {468--469}\BibitemShut {NoStop}%
\bibitem [{\citenamefont {Bickel}\ and\ \citenamefont
  {Doksum}(2015)}]{bickel2015mathematical}%
  \BibitemOpen
  \bibfield  {author} {\bibinfo {author} {\bibfnamefont {P.~J.}\ \bibnamefont
  {Bickel}}\ and\ \bibinfo {author} {\bibfnamefont {K.~A.}\ \bibnamefont
  {Doksum}},\ }\href@noop {} {\emph {\bibinfo {title} {Mathematical statistics:
  basic ideas and selected topics, volume I}}},\ Vol.\ \bibinfo {volume} {117}\
  (\bibinfo  {publisher} {CRC Press},\ \bibinfo {year} {2015})\BibitemShut
  {NoStop}%
\bibitem [{\citenamefont {Montavon}\ \emph {et~al.}(2012)\citenamefont
  {Montavon}, \citenamefont {Orr},\ and\ \citenamefont
  {M{\"u}ller}}]{montavon2012neural}%
  \BibitemOpen
  \bibfield  {author} {\bibinfo {author} {\bibfnamefont {G.}~\bibnamefont
  {Montavon}}, \bibinfo {author} {\bibfnamefont {G.}~\bibnamefont {Orr}},\ and\
  \bibinfo {author} {\bibfnamefont {K.-R.}\ \bibnamefont {M{\"u}ller}},\
  }\bibfield  {title} {\bibinfo {title} {Neural networks-tricks of the trade
  second edition},\ }\href@noop {} {\bibfield  {journal} {\bibinfo  {journal}
  {Springer, DOI}\ }\textbf {\bibinfo {volume} {10}},\ \bibinfo {pages} {978}
  (\bibinfo {year} {2012})}\BibitemShut {NoStop}%
\end{thebibliography}%

\clearpage

\appendix
\section{Conventions for Action}
The form of the action $S$ used in the main text is
\begin{align}
    S(\phi) = \sum_{x \in \Lambda} - 2 \kappa \sum_{\hat{\mu}=1}^2 \phi(x)& \phi(x + \hat{\mu}) + (1 - 2 \lambda) \phi(x)^2 \nonumber \\
&+ \lambda \, \phi(x)^4 \,.
\end{align}
It can be obtained by starting from the (more standard) action
\begin{align}
    S(\varphi) = \sum_{x \in \Lambda} a^2 \frac{1}{2} \sum_{\hat{\mu}=1}^2 &\frac{(\varphi(x+a \hat{\mu}) - \varphi(x))^2}{a^2} \nonumber \\
    &+ \frac{m_0^2}{2} \varphi^2(x) + \frac{g_0}{4!} \varphi^4(x) 
\end{align}
and performing the following re-definitions
\begin{align}
    \varphi &= (2 \kappa)^{\frac{1}{2}} \phi \,, \\
    (a m_0)^2 &= \frac{1 -2 \lambda}{\kappa} - 4 \,, \\
    a^2 g_0 &= \frac{6 \lambda}{\kappa^2} \,.
\end{align}

\section{A brief overview of Deep Learning}
\paragraph{Neural Networks}
Neural networks are a machine learning algorithm which has proven to be particularly powerful. A neural network is build of layers which are defined by
$$
y^{(l)}(x) = \sigma ( W^l x + b^l) \,,
$$
where $x\in \mathbb{R}^n$, $y^l \in \mathbb{R}^m$ are input and output of the layer. The output of the layer is also often called the activation of the layer. The weights $W^l \in \mathbb{R}^{m,n}$ and the bias $b^l \in \mathbb{R}^n$ are the learnable parameters of the neural network. The non-linearity $\sigma: \mathbb{R} \to \mathbb{R}$ is a non-linear function which is applied element-wise to the components of $W^l x + b^l$. Widely-used activation function are $\sigma(x)=\max(x,0)$ or $\sigma(x)=\tanh(x)$.

A neural network consists of $L$ such layers, i.e.
$$
    g(x) = (y^{(L)} \circ \dots \circ y^{(1)})(x) \,.
$$
It is important to note that the weights $W^l$ and biases $b^l$ do not have to be of the same dimensionality for each layer (although we did not make this explicit in our notation). It is also important to note that we merely described the most simple type of neural network, namely a fully-connected neural network. There is a zoo of other neural networks but we will refrain from a more detailed discussion as it is not needed for our purposes (see \cite{goodfellow2016deep} for an overview).

\paragraph{Learning Parameters with Backpropagation}
The parameters of the neural networks,
\begin{align*}
\mathcal{W} = \{( W^l, \, b^l ) \,, \;\; i = 1,\dots,L \} \,,
\end{align*}
are determined by minimizing a certain loss function $\mathcal{L}$ by gradient descent (see \eqref{eq:varfreeenergy} for the particular loss function used in this work). It is important to emphasize that the number of parameters are typically large (of order $10^3-10^6$ for typical modern neural networks).  
It is therefore clear that one cannot determine the gradient by finite-difference (as we would need calculate the finite difference ratio for each of these parameters which is prohibitively expensive). 

The basic idea for calculating the gradient $\nabla_\mathcal{W} \mathcal{L}$ is to use the fact that we know the functional form of the neural network: the gradient of the loss is given by
\begin{align*}
\nabla_\mathcal{W} \mathcal{L} = \frac{\partial \mathcal{L}}{\partial y^L}\frac{\partial y^L}{\partial y^{L-1}} \dots \frac{\partial y^1}{ \partial x} \,.
\end{align*}
Each term in this expression is known, for example
\begin{align*}
\frac{\partial y^{l+1}}{ \partial y^l} = \sigma'(y^l) W^l \,.
\end{align*}
For a fixed non-linearity, we know the analytical form of the derivative $\sigma'$. This observation leads to the following algorithm: we first perform a \emph{forward pass} of the neural network, i.e. starting from the input $x$, we calculate the activations $y^l$ for each layer and store them in memory. This process ends with the final activation $y^L$ which is, by definition, the output of the neural network. The gradient of each layer 
$\frac{\partial y^{l+1}}{ \partial y^l}=\sigma'(y^l) W^l$  can then directly be calculated (as we have stored the activation $y^l$). Crucially, we only need the matrix product of these Jacobians and it is efficient to start by calculating the gradient with respect to the output layer $L$, then the layer $L-1$ and so forth. This is because the loss function has a scalar output value and therefore the matrix product of the Jacobians 
\begin{align*}
\frac{\partial \mathcal{L}}{\partial y^L}\frac{\partial y^L}{\partial y^{L-1}} \dots \frac{\partial y^{l+1}}{ \partial y^{l}}
\end{align*}
is a vector with the same number of components as $y^l$. We can therefore save memory by simply overwriting the stored activation $y^l$. This algorithm is called \emph{backpropagation} and allows us to calculate the gradient $\nabla_{\mathcal{W}} \mathcal{L}$ for roughly the same cost as a forward pass of the neural network.

\section{Relation between $\textrm{Var}(C)$ and KL divergence}
\begin{figure}[tb]
    \centering
    \includegraphics[width=0.5\textwidth]{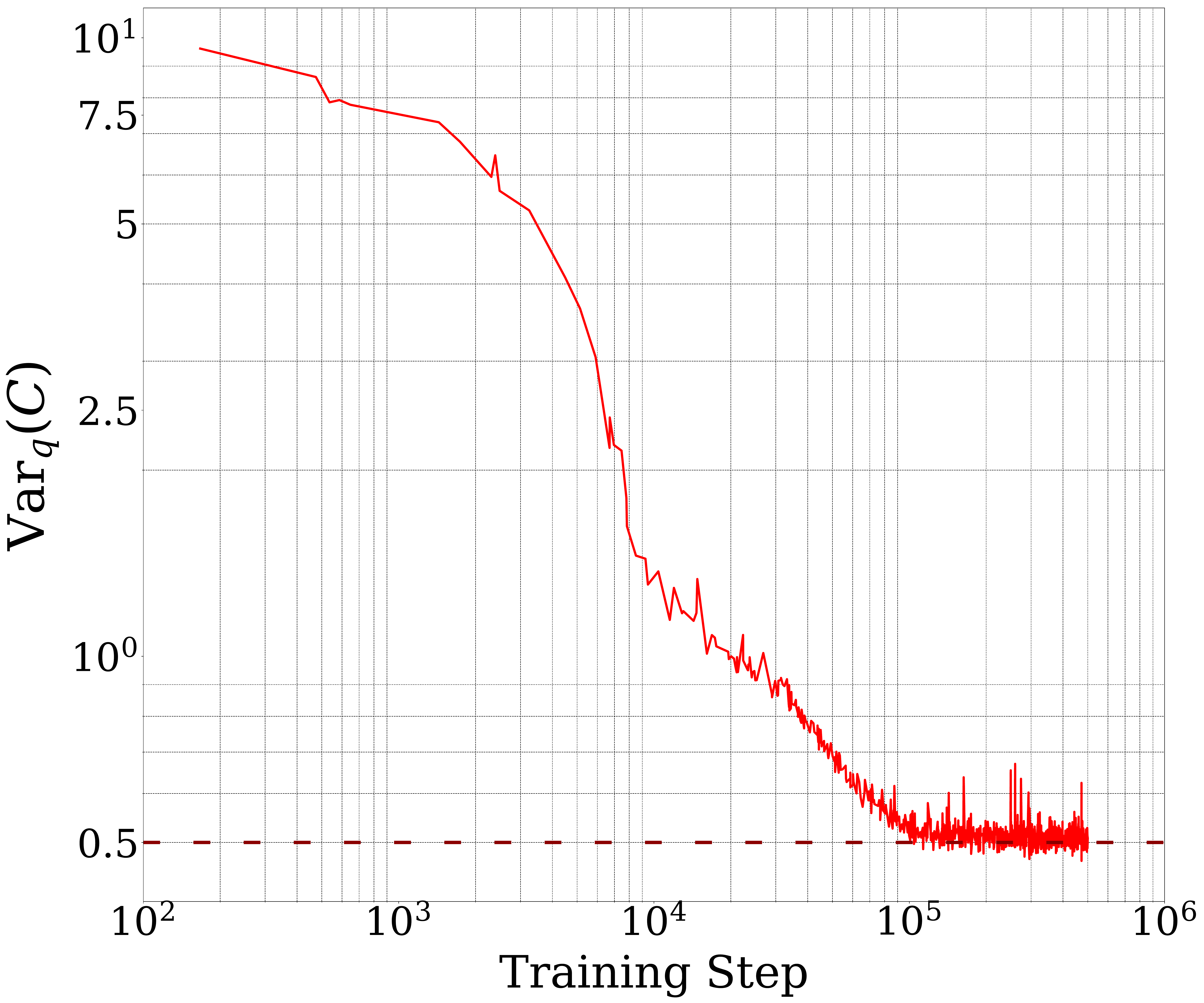}
    \caption{The variance $\textrm{Var}_q(C)$ decreases during training. We use hopping parameter $\kappa=0.3$ and bare coupling $\lambda=0.022$ for a $16 \times 8$ lattice.} 
    \label{fig:convergence}
\end{figure}

\begin{theorem}
Let $C(\phi) = S(\phi) + \ln q(\phi)$. The following relation between the KL divergence and the variance of $C$ holds: 
\begin{align*}
    \textrm{KL} (q_\theta || p) = \tfrac{1}{2} \, \textrm{Var}_q(C) +  \mathcal{O}(\mathbb{E}_q [|w - 1|^3]) \,,
\end{align*}
where $w(\phi)=\frac{p(\phi)}{q(\phi)}$ is the normalized importance weight.
\begin{proof}
The expectation value of the normalized importance weight is
\begin{align*}
    \mathbb{E}_q \left[ \frac{p}{q}\right] = \int \mathcal{D}[\phi] \, p(\phi) = 1 \,.
\end{align*}
The Kullback-Leibler divergence can be rewritten in terms of the normalized importance weight 
\begin{align*}
    \textrm{KL}(q | p) = \mathbb{E}_q\left[ \ln \frac{q}{p}\right] = - \mathbb{E}_q \left[ \ln w \right] \,. 
\end{align*}
We now expand the KL divergence around the expectation value of the normalized importance weight
\begin{align*}
    \textrm{KL}( q | p) &= - \mathbb{E}_q \left[ \ln \left(1 - (w - 1)\right) \right] \\
    &=  \mathbb{E}_q \left[ \sum_{j=0}^\infty \frac{(-1)^j}{j} (w-1)^j \right] \\
    &= \underbrace{\mathbb{E}_q[w - 1]}_{=0} + \frac{1}{2} \mathbb{E}_q [(w - 1)^2] + \mathcal{O}(\mathbb{E}_q [ |w-1|^3]) \,. 
\end{align*}
We now relate this expression to the variance of $C$. To this end, we first observe that $C=-\ln \tilde{w}$, where $\tilde{w}=\exp(-S) /q $ is the unnormalized importance weight. We then rewrite the expectation value of $C$ as
\begin{align*}
    \mathbb{E}_q [C] &= - \mathbb{E}_q [ \ln \tilde{w}] \\
    &= - \mathbb{E}_q \left[ \ln w + \ln Z \right] \\
    &= - \ln Z - \mathbb{E}_q [\ln w] \\
    &= - \ln Z + \textrm{KL}(q | p) \\
    &= - \ln Z + \mathcal{O}(\mathbb{E}_q[(w-1)^2]) \,,
\end{align*}
where the last step uses the expansion for the KL divergence derived above.
It then follows its variance is given by
\begin{align*}
    \textrm{Var}_q(C) &= \mathbb{E}_q\left[ \left( C - \mathbb{E}_q [C] \right)^2\right] \\
    &= \mathbb{E}_q \left[ \left( - \ln \tilde{w} + \mathbb{E}_q [\ln \tilde{w}] \right)^2 \right] \\
    &=  \mathbb{E}_q \Bigg[ \Big( \underbrace{- \ln \tilde{w} + \ln Z}_{=-\ln w} + \mathcal{O}\left(\mathbb{E}_q[(w-1)^2]\right) \Big)^2 \Bigg]
\end{align*}
Expanding the logarithm around $\mathbb{E}_q[w]=1$ again, we obtain
\begin{align*}
    \textrm{Var}_q[C] = \mathbb{E}_q [(w-1)^2] + \mathcal{O}\left( \mathbb{E}_q [|w-1|^3] \right) \,.
\end{align*}
Combining this expression with the expansion derived for the KL divergence, we obtain the claim of the theorem.
\end{proof}
\end{theorem}
The higher-order moments will be small towards the end of the training process for which $q \approx p$ and thus $w \approx 1$. Thus, the variance of $C$ will become small. We indeed observe this behaviour in our numerical experiments, see Figure~\ref{fig:convergence} for an example.

\section{Analytic Solution for Partition Function}
\begin{figure*}[ht]
    \centering
    \includegraphics[width=\textwidth]{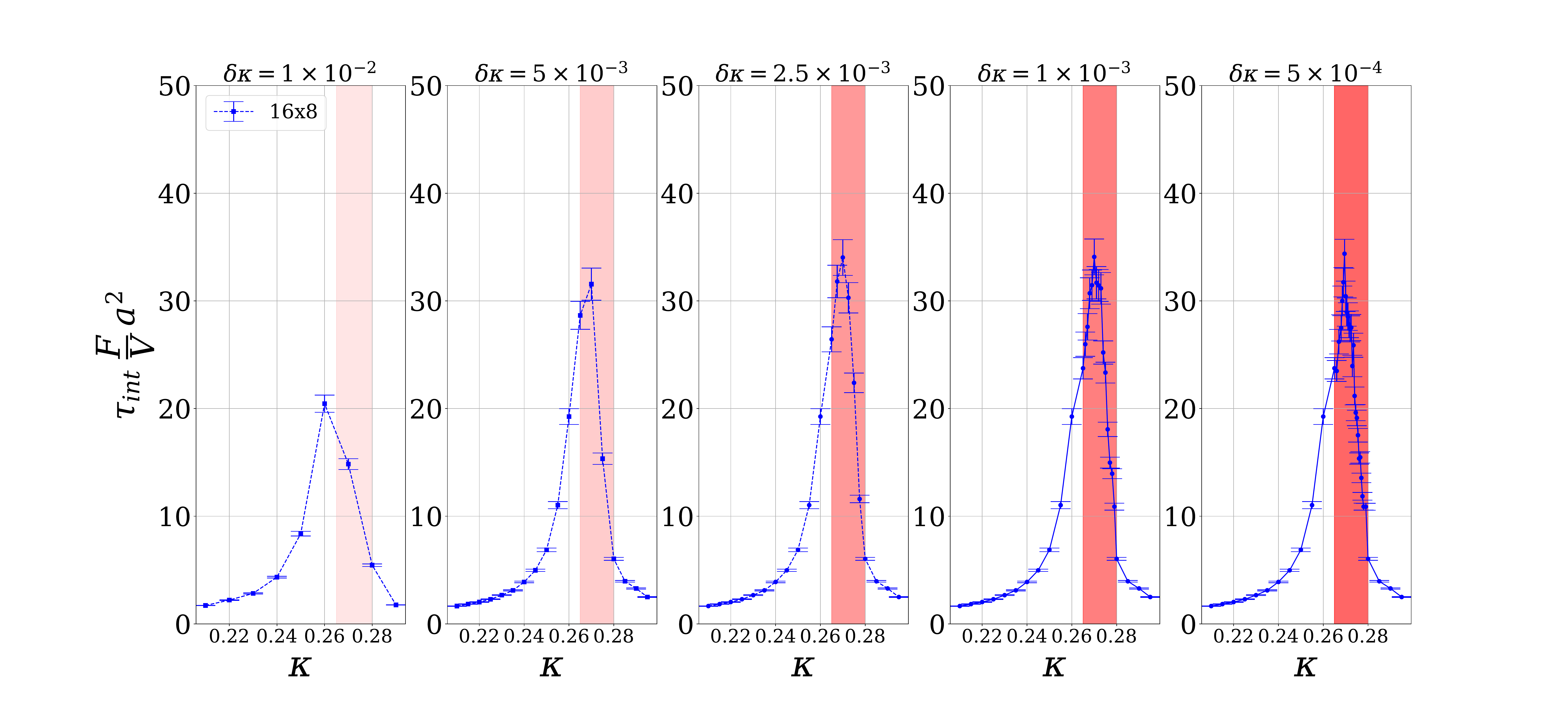}
    \caption[width=\textwidth]{Integrated autocorrelation time of the free energy during refinement of the step size. The shaded red areas refers to the interval in hopping parameter $\kappa\in [0.265, 0.28]$ for which the refinement is applied. Darker shading indicates narrower step size. The experiments were performed using the overrelaxed HCM algorithm.}
    \label{fig:autocorrelation}
\end{figure*}

The action of the scalar field theory is given by
\begin{align*}
S = \sum_{x \in \Lambda} - 2 \kappa \sum_{\hat{\mu}=1}^2 \varphi(x) \varphi(x + \hat{\mu}) + (1 - 2 \lambda) \varphi(x)^2
+ \lambda \, \varphi(x)^4 \,.
\end{align*}
We want to calculate the partion function
\begin{align*}
    Z = \int \mathcal{D} [\phi] \, \exp(-S(\phi)) \,,
\end{align*}
which for vanishing hopping parameter $\kappa$ decouples in independent integrals of each lattice site of the lattice $\Lambda$:
\begin{align*}
    Z = \prod_{x \in \Lambda} \left( \int \text{d}\phi(x) \, \exp(- \lambda \, \phi(x)^4 - (1 - 2 \lambda) \, \phi(x)^2 ) \right)
\end{align*}

The partition function can then be calculated analytically using the integral
\begin{align*}
    \int \exp(- a x^4 - b x^2) \, \textrm{d}x = \sqrt{\frac{b}{4a}} \, \exp\left(\frac{b^2}{8 a}\right) \, K_{\frac{1}{4}}\left( \frac{b^2}{8a} \right)  \,,
\end{align*}
where $K_n$ is the modified Bessel function of the second kind.
Using this formula, we obtain the following analytic form of the free energy
\begin{align}
    F = - T \, |\Lambda| \, \ln (z) \,, \nonumber
\end{align}
where we have defined
\begin{align*}
z(\lambda) = \sqrt{\frac{1-2\lambda}{4 \lambda}} \, \exp\left(\frac{(1-2 \lambda)^2}{8 \lambda}\right) \, K_{\frac{1}{4}}\left( \frac{(1-2 \lambda)^2}{8 \lambda} \right) \,.
\end{align*}
This result corresponds to the zeroth order of the hopping expansion \cite{gattringer2009quantum} of the partition function $Z$ and one may, in principle, also calculate higher-order corrections. However, they are not needed for our purposes.

\section{Error Analysis for Free Energy Estimator}
Our discussion is based on \cite{nowozin2018debiasing} which discussed the same results in the context of variational inference. We provide a review here since these results may be hard to extract for physicist not familiar with variational inference. We also point out a subtlety that was not discussed in the previous work.

The estimator for the free energy is given by
\begin{align}
    \hat{F} = - T \, \ln \frac{1}{N} \sum_{i=0}^N \tilde{w} (\phi_i)\,, && \phi_i \sim q_\theta \,. 
\end{align}
Theorems for the variance and bias of this estimator are discussed in the following. 
For this, we use the \emph{delta method of moments} which is summarized in the following theorem.

\begin{theorem}
Let $\hat{X}_N=\frac{1}{N}\sum_{i=1}^N X_i$ be the sample mean of independent and identically distributed random variables $X_i$ with $\mathbb{E}\left[X_i^{2k+2}\right]<\infty$ for $k\in\{0,1\}$. Let $h$ be a real-valued function with uniformly bounded derivatives. 
It then holds that
\begin{align*}
    \mathbb{E}\left[ h(\hat{X}_N) \right] = c_0 + \frac{c_1}{N} + \mathcal{O}\left(\frac{1}{N^2}\right) \,,
\end{align*}
where 
\begin{align*}
    c_0 = h(\mu)\,, && c_1 = h''(\mu) \frac{\sigma^2}{2} \,, 
\end{align*}
with $\sigma^2 = \mathbb{E} \left[ (X - \mathbb{E}X)^2 \right]$ and $\mu=\mathbb{E}\left[X\right]$.
\end{theorem}
We refer to Chapter~5.3 of \cite{bickel2015mathematical} for a proof and more details.

The application of the delta method to the free energy estimator $\hat{F}$ is, in practise, subject to a subtlety regarding the bounded differentiablity of the function $h$. We will ignore this subtlety in the following and return to it at the end of the section.

\begin{theorem}
The bias of $\hat{F}$ is given by
\begin{align}
    \mathbb{B}[-\beta \hat{F}] = -\frac{1}{2 \, N} \frac{\mathbb{E}_q \left[ ( \tilde{w} - \mathbb{E}_q[\tilde{w}] )^2\right]}{\mathbb{E}_q [\tilde{w}]^2} + \mathcal{O}(N^{-2}) \,, \label{eq:app_bias_est}
\end{align}
assuming that $\mathbb{E}_q[\tilde{w}^{2k+2}] < \infty$ for $k=0,1$.

\begin{proof}
The bias of $-\beta \hat{F} = \ln \hat{Z}$ is given by
\begin{align*}
     \mathbb{B}[-\beta \hat{F}] = \mathbb{E}_q [ \ln \hat{Z}] - \ln Z \,. 
\end{align*}
Using the delta method for moments, we derive that
\begin{align}
 \mathbb{E}_q [ \ln \hat{Z}] = \ln Z - \frac{1}{2 N Z^2} \mathbb{E}_q \left[ (\tilde{w} - \mathbb{E}_q[\tilde{w}])^2\right] + \mathcal{O}\left(N^{-2}\right) \,,  \label{eq:biasexpansion}
\end{align}
where we have used that $h(x)=\ln(x)$ has second derivative $h''(x)=-\frac{1}{x^2}$. The proof then concludes by observing that
\begin{align*}
    \mathbb{E}_q \left[ \tilde{w} \right] = Z \,.
\end{align*}
\end{proof}
\end{theorem}

\begin{theorem}
The variance of $\hat{F}$ is given by
\begin{align} 
\textrm{Var}\left[-\beta \hat{F}\right] = \frac{1}{N} \frac{\mathbb{E}_q\left( \tilde{w} - \mathbb{E}_q\left[\tilde{w} \right]\right)^2}{(\mathbb{E}_q[\tilde{w}])^2} + \mathcal{O}\left(\frac{1}{N^2}\right) \,, \label{eq:app_var_est}
\end{align}
assuming that $\mathbb{E}_q[\tilde{w}^{2k+2}] < \infty$ for $k=0,1$.
\begin{proof}
The variance can be written as
\begin{align*}
    &\textrm{Var}[-\beta \hat{F}] = \mathbb{E}_q [ (\ln \hat{Z})^2 ] - \mathbb{E}_q[\ln \hat{Z}]^2 \,.
\end{align*}
We now evaluate both terms on the right-hand-side individually using the delta method. For the first term, we use the delta method with $h(x)=(\ln x)^2$ which has second derivative
\begin{align*}
    h''(x) = \frac{2}{x^2} - 2 \frac{\ln(x)}{x^2} \,. 
\end{align*}
Using this expression, we then obtain that $\mathbb{E}_q [ (\ln \hat{Z})^2 ]$ is equal to
\begin{align*}
    (\ln Z)^2 + \left( \frac{1}{Z^2}  - \frac{\log Z}{Z^2}\right) \frac{\mathbb{E}_q\left[ (\tilde{w} - \mathbb{E}_q [\tilde{w}])^2\right]}{N} + \mathcal{O}(N^{-2}) 
\end{align*}
For the squared expectation value, we use the expansion \eqref{eq:biasexpansion} derived in the proof for the bias. This gives that $(\mathbb{E}_q \ln \hat{Z})^2$ is equal to
\begin{align*}
    &\left( \ln Z - \frac{1}{2 N Z^2} \mathbb{E}_q \left[ (\tilde{w} - \mathbb{E}_q[\tilde{w}])^2\right] + \mathcal{O}\left(N^{-2}\right) \right)^2 \\
    &= (\ln Z)^2 - \frac{1}{N Z^2} \mathbb{E}_q \left[ (\tilde{w} - \mathbb{E}_q[\tilde{w}])^2\right] + \mathcal{O}\left(N^{-2}\right) \,.
\end{align*}
Subtracting these two expressions, it then follows
\begin{align*}
  \textrm{Var}[-\beta \hat{F}] =   \frac{1}{Z^2} \frac{\mathbb{E}_q\left[ (\tilde{w} - \mathbb{E}_q [\tilde{w}])^2\right]}{N} + \mathcal{O}(N^{-2}) \,,
\end{align*}
and the proof concludes by observing that $Z = \mathbb{E}_q [\tilde{w}]$.
\end{proof}
\end{theorem}
A few remarks are in order: from the theorems, it follows that the standard deviation of the estimator $\hat{F}$ is of order $\mathcal{O}(1/\sqrt{N})$. In the large $N$ limit, we can therefore neglect the bias correction as it is of order $\mathcal{O}(N^{-1})$.
Furthermore, we can replace the expectation values in the theorems by the sample mean up to (negligible) higher-order corrections. In practise, we therefore use these results to estimate the variance and bias of $\hat{F}$. Alternatively, one can use a standard Jackknife analysis to estimate variance and bias (see for example \cite{gattringer2009quantum}). In our experiments, we use both methods to estimate the errors and check that they lead to consistent results. Lastly, we remark that error estimators for general observables involving the partition function can be derived, see \cite{nicoli2020asymptotically}.

As mentioned above, the delta method requires that the derivatives of the function $h$ are (uniformly) bounded. For a generic LQFT, this will not be the case for $h(x)=\ln(x)$ since its derivatives diverge for $x \to 0^+$. To the best of our knowledge, the same problem will generically arise in the context of variational inference but seems to have not been discussed in the literature. 

To address this subtlety, one could require that the action of the lattice quantum field theory is bounded. For example, this can be ensured by putting the field theory in a box potential. Since only very high energy configurations are affected by this (for suitably large choice of the box potential) and since these configurations are extremely unlikely to be sampled, this modification will have no practical effect on the numerical experiments. After this modification, $\hat{Z}$ is bounded from below and $h^{(n)}(\hat{Z})$ is also bounded as a result.

More rigorously, the result for the variance can be derived without assumptions on a bound for the derivatives by using the \emph{delta method for in law approximation} which takes the following form
\begin{theorem}
Let $\hat{X}_N=\frac{1}{N}\sum_{i=1}^N X_i$ be the sample mean of independent and identically distributed random variables $X_i$ with $\mathbb{E}\left[X_i^k\right]<\infty$ for $k\in\{1,2\}$. Let $h$ be a differential function at $\mu=\mathbb{E}_q[X]$. Then
\begin{align*}
   \sqrt{n} \left(h(\hat{X}_N) - h(\mu)\right) \overset{D}{\to} \mathcal{N}(0, \sigma^2(h) ) \,,
\end{align*}
where $\sigma^2(h)=h'(\mu) \textrm{Var}(X)$.
\end{theorem}
For a proof, we again refer to \cite{bickel2015mathematical}, see Theorem~5.3.3. Applying this theorem to the free energy estimator $-\beta \hat{F} = \ln \hat{Z}$, we obtain the same expression for its variance as derived above. However, the theorem does not require any bound on the derivatives of $h(x)=\ln(x)$.

\section{Systematic Errors}

\begin{figure}[ht]
    \centering
    \includegraphics[width=0.4\textwidth]{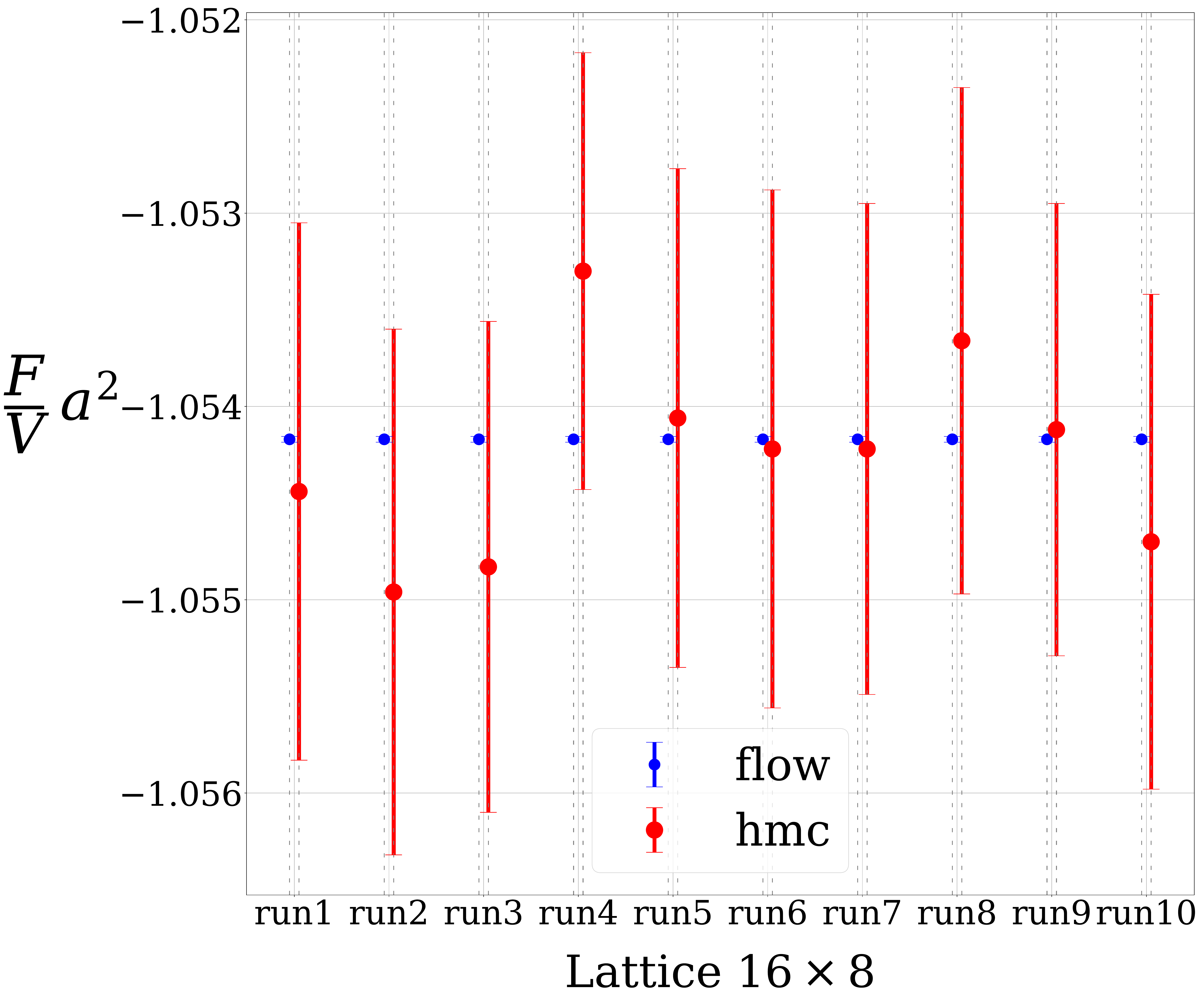}
    \caption{Free energy estimation as for Figure~\ref{fig:freeeneergy} repeated ten times for the $16\times8$ lattice.}
    \label{fig:repeatedruns}
\end{figure}

In this section, we discuss the various sources of systematic error relevant for both the flow-based and MCMC-based estimation methods and discuss how they can be assessed. 
\\
\begin{figure}[ht]
    \centering
    \includegraphics[width=0.4\textwidth]{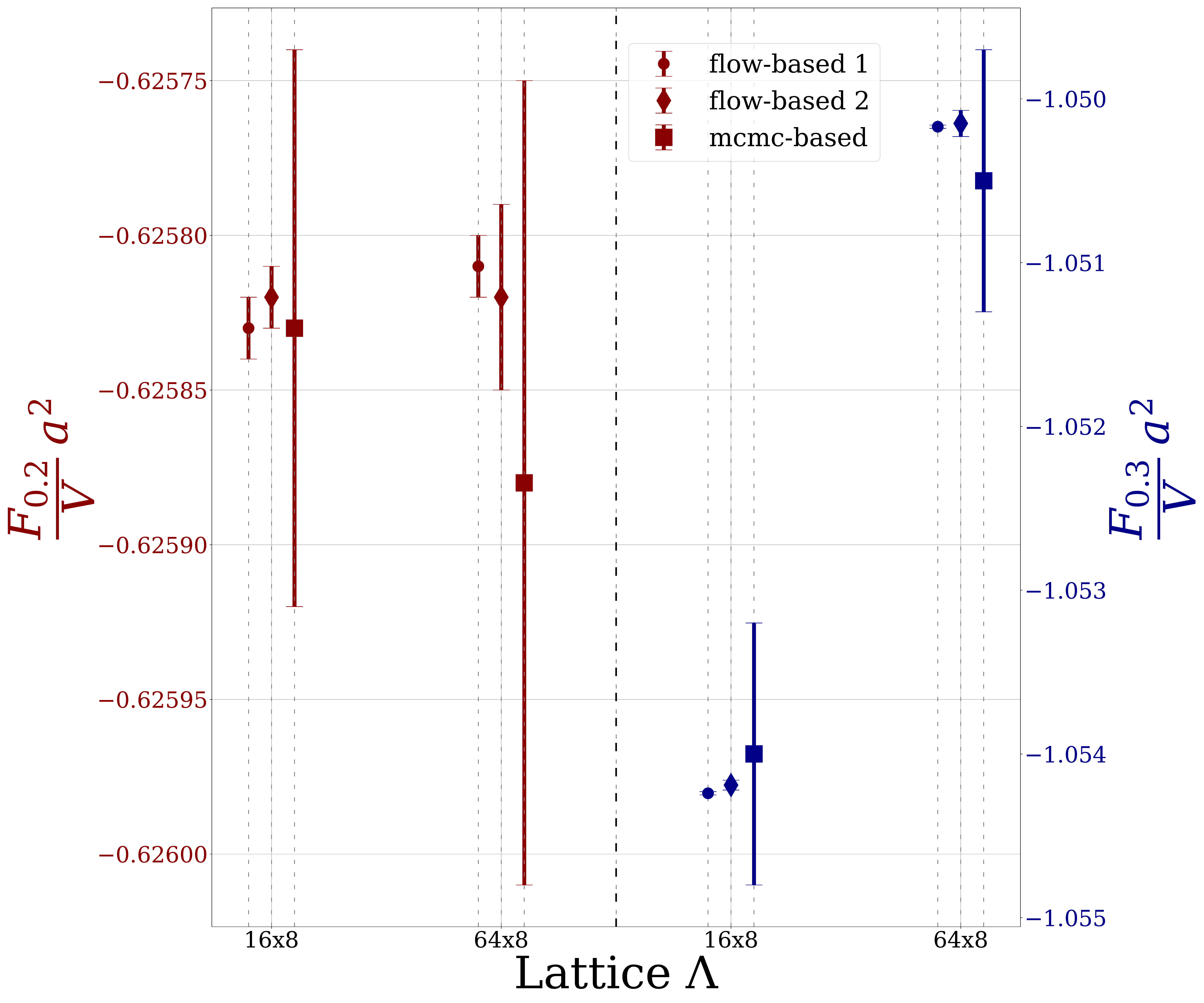}
    \caption{The left-hand and right-hand sides show the estimates for the free energy at hopping parameters $\kappa=0.2$ and $\kappa=0.3$ for fixed bare coupling $\lambda=0.022$ respectively. We used the same procedure as for Figure~\ref{fig:freeeneergy} on both the smallest and largest lattice for illustration. The first and second flow-based method use as a sampler $p$ and $q$ respectively and lead to compatible results. We use the uwerr method for MCMC error analysis.}
    \label{fig:mode_dropping}
\end{figure}
\paragraph{Mode-dropping for Generative Model:}
We ensure that no mode-dropping for the generative model takes place, i.e. all modes of the target distribution $p$ are captured by the generative model $q$. Mode-dropping can cause underestimation of the partition function $Z$. To this end, we use the relation
\begin{align*}
 \mathbb{E}_{\phi \sim p} [\tilde{w}^{-1}(\phi)] = \frac{1}{Z} \int \mathcal{D}[\phi] \, e^{-S(\phi)} \, \frac{q(\phi)}{e^{-S(\phi)}} = Z^{-1} \,.
\end{align*}
The left-hand-side can be estimated using a single Markov chain. Note that we sample from the target distribution $p$ as opposed to the generative model $q$. Therefore, the  estimator is consistent even if $q$ is mode-dropping. The resulting estimate $\hat{Z}$ for the partition function can then be plugged into \eqref{eq:freeenergyest} to obtain an estimate for the free energy. It is then checked that the result is consistent with the one obtained from \eqref{eq:partition_est}, see Figure~\ref{fig:mode_dropping}. We stress that this consistency check only requires a single Markov chain and is therefore equal in cost to the overlap consistency check of the MCMC method discussed below.
\paragraph{Bias due to imperfect training:} From the theoretical analysis of the variance of the estimator \eqref{eq:app_var_est}, it is expected that it has a standard deviation with the typical $N^{-\frac{1}{2}}$ fall-off in the number of samples $N$. On the other hand, the bias of the estimator \eqref{eq:app_bias_est} due to imperfect training has a subleading $N^{-1}$ fall-off. We check carefully that our error estimates indeed show the theoretically predicted $N^{-\frac12}$ behaviour in our numerical experiments.
\paragraph{Repeated runs:} We repeated the estimate for the smallest lattice ten times. The resulting estimates are shown in Figure~\ref{fig:repeatedruns}. The sample standard deviation is consistent with our error estimates. Furthermore, the MCMC-based method does not systematically over- or underestimate with respect to the flow. 

\begin{figure}[h]
    \centering
    \includegraphics[width=0.4\textwidth]{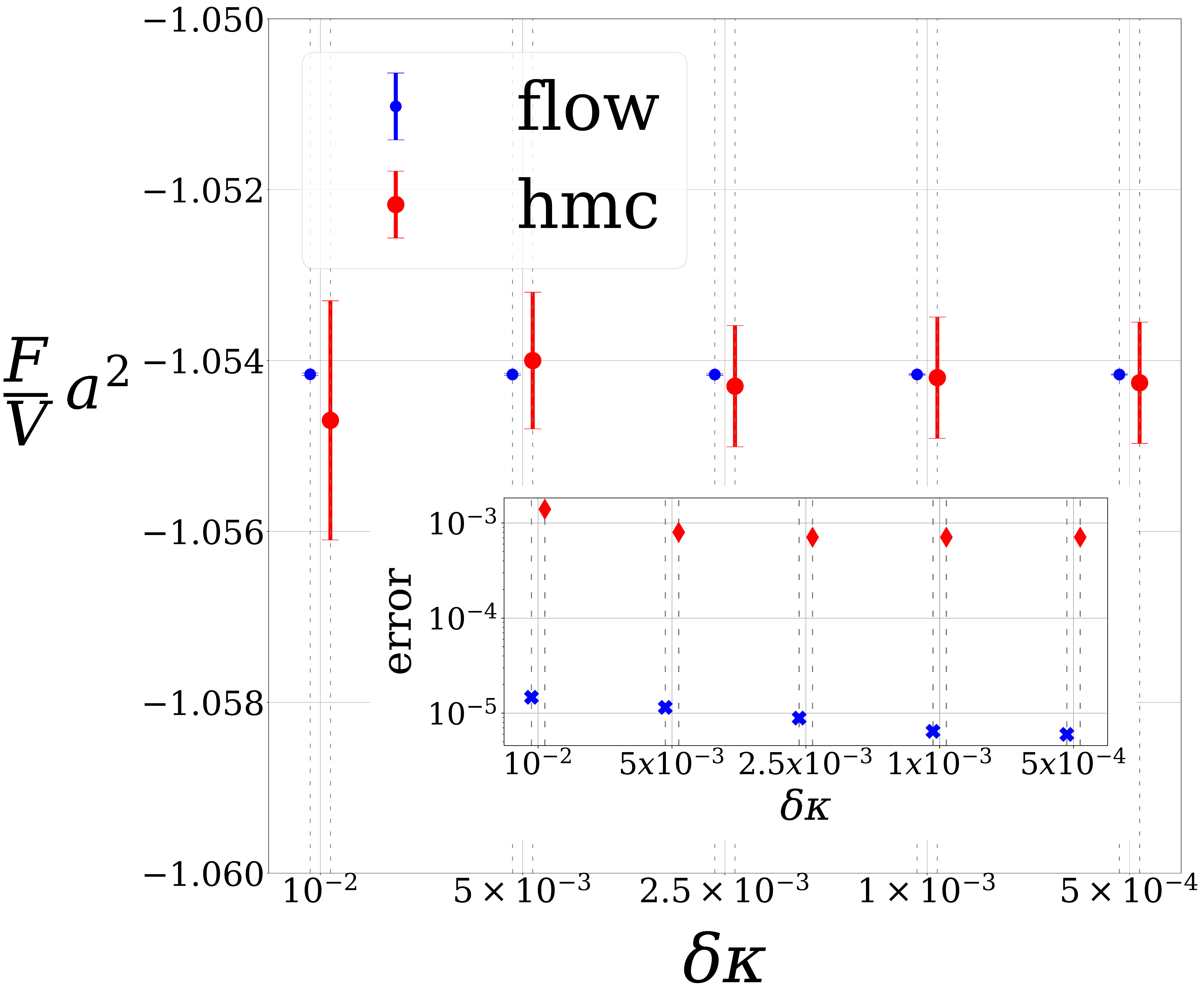}
    \caption{Free energy at $\kappa=0.3$ obtained by both MCMC-based and flow-based method. For the MCMC-based method, different step size $\delta \kappa$ of the hopping parameter were used. Details on the step sizes $\delta \kappa$ are summarized in Table~\ref{tab:stepsize}. For the flow-based method, we use the same number of samples as for the corresponding refined MCMC method. As a result, also the error of the flow's estimate decreases.}
    \label{fig:stepsize}
\end{figure}

\begin{figure}[h]
    \centering
    \includegraphics[width=0.4\textwidth]{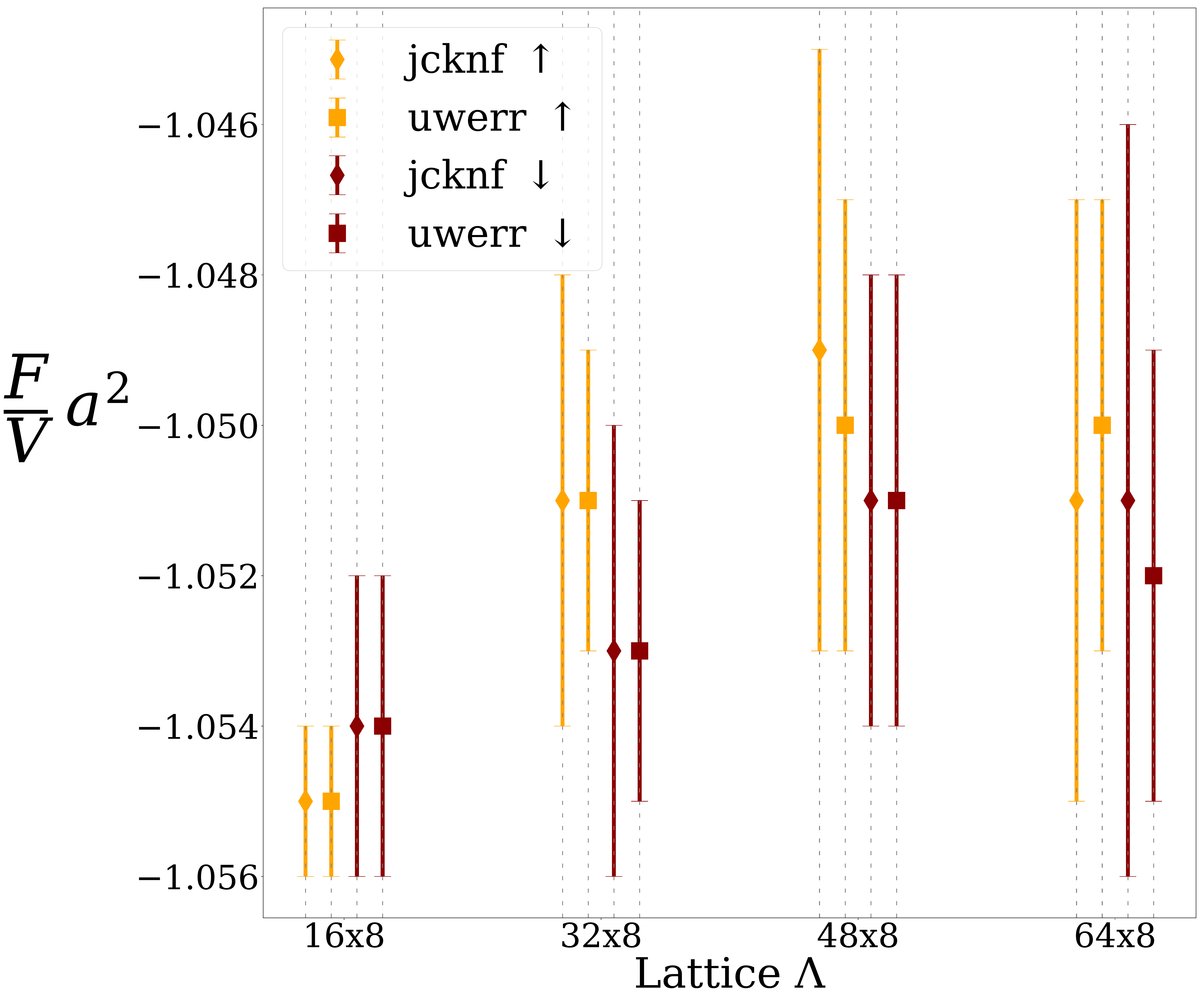}
    \caption{Free energy at $\kappa=0.3$ and $\lambda=0.022$ using both $Z_i/Z_{i+1}$ (down) and $Z_{i+1}/Z_i$ (up). We use the same setup (i.e. number of steps, hopping parameter change $\delta \kappa$, etc) as for Figure~\ref{fig:freeeneergy}.}
    \label{fig:freeenergyupdown}
\end{figure}
\paragraph{Step size for MCMC:} As explained in the main text, the free energy difference $\Delta F_{e\,b}$ is calculated in steps
\begin{align*}
    \Delta F_{e \, b} = \Delta F_{e, i_K} + \Delta F_{i_K \, i_{K-1}} + \dots + \Delta F_{i_1 \, b} \,.
\end{align*}
In the following, we will analyze the dependency of our results on the chosen steps.

We start from an initial step size corresponding to a change in hopping parameter $\kappa$ of $\delta \kappa = 0.05$. Between $\kappa=0.2$ and $\kappa=0.3$, we however take a finer step size of $\delta \kappa = 0.01$. Since we are interested in the free energy difference $\Delta F_{e \, b}$ between $\kappa_b=0.0$ and $\kappa_e=0.3$, this corresponds to running $K=14$ Markov chains. We focus on the $16 \times 8$ lattice and use an overelaxed HMC algorithm to sample $400k$ configurations for each chain. The overrelaxation is performed every $10$ steps.   

We then repeatedly refine the step size in a certain subregion around the critical $\kappa$ value. The details can be found in Table~\ref{tab:stepsize}. 

The results of this analysis are shown in Figure~\ref{fig:stepsize}. We observe that the error of the estimator does not significantly decrease. We note that the error of the flow decreases during refinement because its free energy estimation uses the same number of samples as all Markov chains combined (and this number increases by the additional refinement steps). 

\paragraph{Mode-dropping for MCMC:}

In order to ensure that the distributions $p_i$ and $p_{i+1}$ in
\begin{align*}
   \mathbb{E}_{p_i} \left[ \frac{\exp(-S_{i+1})}{\exp(-S_i)} \right]= \frac{1}{Z_i} \int \mathcal{D}[\phi] \, e^{-S_{i}(\phi)} \,\frac{e^{-S_{i+1}(\phi)}}{e^{-S_i(\phi)}} = \frac{Z_{i+1}}{Z_i} \,
\end{align*}
have sufficient overlap, we also estimate $\frac{Z_i}{Z_{i+1}}$ by exchanging $p_i$ with $p_{i+1}$ in the relation above. We then check that this leads to compatible results, see Figure~\ref{fig:freeenergyupdown}. We note that this consistency check is relatively cheap as it requires running one additional Markov chain.
We also study the dependence of the integrated autocorrelation of the free energy on the refinement of $\delta \kappa$, see Figure~\ref{fig:autocorrelation}.

\begin{table}
\caption{\label{tab:stepsize} Details on the refinement analysis. In each refinement stage, we take smaller steps $\delta \kappa$ in a certain subregion of the hopping parameter $\kappa$ trajectory (see last column). The step size taken in this region is shown in the first column. Outside of the most refined region, the same step sizes as in the previous refinement stage are taken. Since each chain is taken to be of the same length, the total number of samples (third column) grows proportional to the number of chains (second column).}
\begin{ruledtabular}
\begin{tabular}{lccr}
$\delta \kappa$ & $\#$ chains & $\#$ samples & refined $\kappa$ region\\
\hline
$0.01$ & 14 & 5.6 M & 0.20-0.30\\
$0.005$ & 24 & 9.6 M &  0.20-0.30\\
$0.0025$ & 40 & 16 M & 0.22-0.30\\
$0.001$ & 76 & 30.4 M & 0.24-0.30\\
$0.0005$ & 88 & 35.2 M & 0.267-0.279\\
\end{tabular}
\end{ruledtabular}
\end{table}

\section{Additional point in parameter space}
\begin{figure}[ht]
    \centering
    \includegraphics[width=0.4\textwidth]{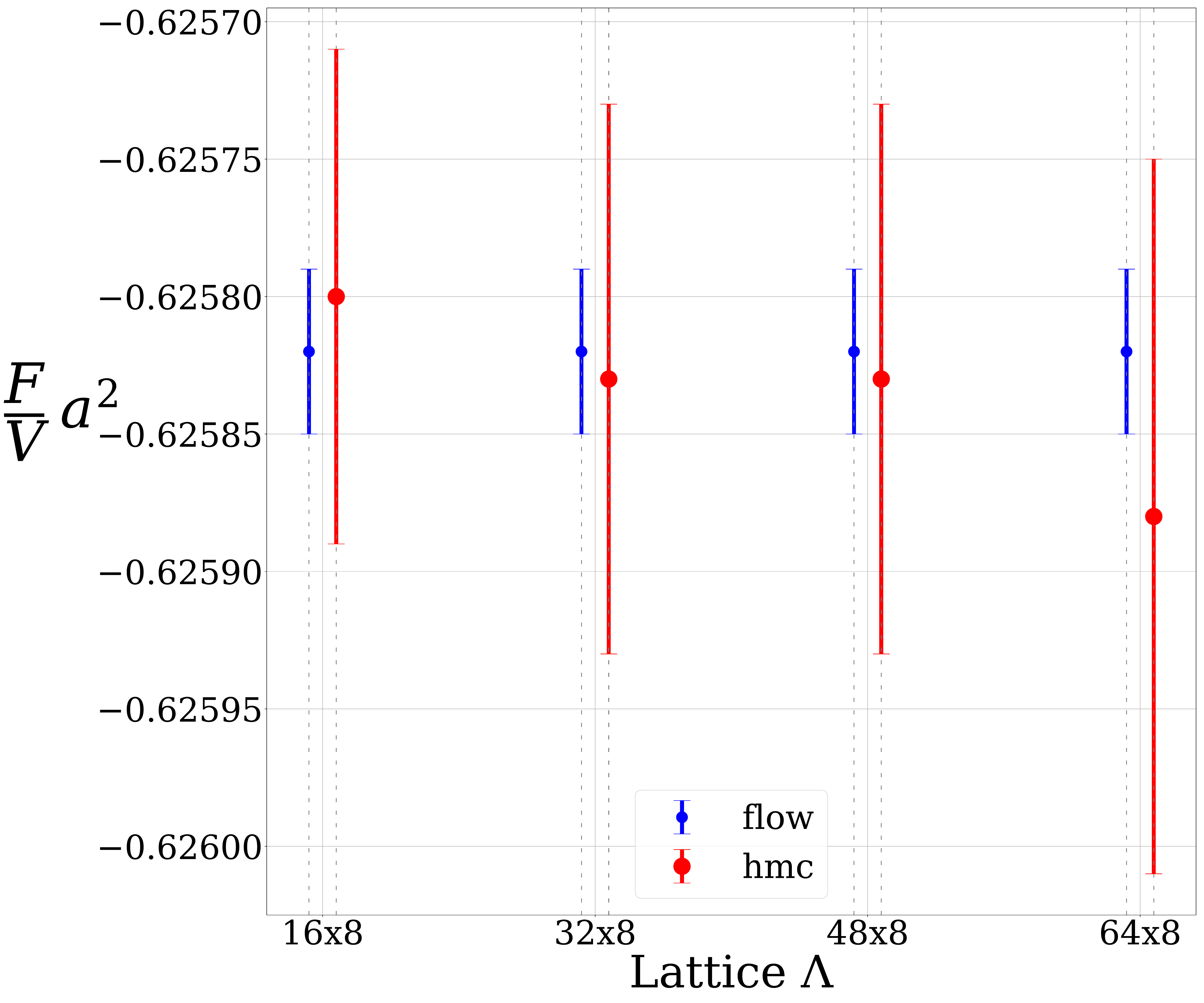}
    \caption{Free energy estimation using the same setup as for Figure~\ref{fig:freeeneergy} but at $\kappa=0.2$.}
    \label{fig:freeenergyapp2}
\end{figure}
In the main text, we demonstrated our method in a case where the MCMC-based method had to cross the critical region to calculate the absolute value of the free energy because we consider this one of the main applications of our method. Since this leads to large integrated autocorrelation times for the MCMC-based method, it has errors which are significantly larger than the generative model estimate.
In the following however, we will focus on a point in phase space for which the MCMC approach does not need to cross a critical region. Namely, we use the same value for the bare coupling $\lambda=0.022$ as in the main text but set the hopping parameter $\kappa=0.2$. As can be seen from Figure~\ref{fig:absmag}, this corresponds to a point before the critical regime. The resulting estimates for the free energy are shown in Figure~\ref{fig:freeenergyapp2}. The MCMC-based estimate has a variance of the same order of magnitude as the flow-based one in this regime.

\section{Details on Numerical Experiments}

\paragraph{HMC:}

We use a HMC algorithm with overrelaxation. Each Markov chain has 5k thermalization steps followed by 400k estimation steps. The sign of the field configuration is flipped every ten steps. 
\paragraph{Training of flow:} For every lattice, we use a normalizing flow with six coupling layers. Each coupling layer \eqref{eq:couplinglayer} has neural network $m$ with five fully-connected layers with no bias and Tanh non-linearities. The hidden layers of $m$ consist of 1000 neurons each. For each coupling layer, we split the input in half to obtain $y^{(u)}$ and $y^{(d)}$, see \eqref{eq:couplinglayer}, using a checkerboard-type partitioning. Consecutive coupling blocks use alternating checkerboard partioning in order to ensure that all lattice sites are updated. We train the flow for 1M steps using an 8k mini-batch. We use ReduceLROnPlateau learning rate scheduler of PyTorch with an initial learning rate of $5 \times 10^{-4}$ and patience of 3k steps. The minimum learning rate was set to $1\times10^{-7}$.
\paragraph{Estimation:}As described in the main text, for HMC-based estimation we use a step size of $\delta \kappa =0.01$ for $\kappa \in [0.2,0.3]$ and a step size of $\delta \kappa=0.05$ for all other values of the hopping parameter. As a result 14 Markov chains are run. In total, the HMC-based method therefore uses $14 \times 400k = 5.6M$ configurations. 
We use the same number of samples for the flow-based estimation. For efficiency, we sample these configurations in mini-batches of 3k samples. 

\begin{figure}[ht]
    \centering
    \includegraphics[width=0.4\textwidth]{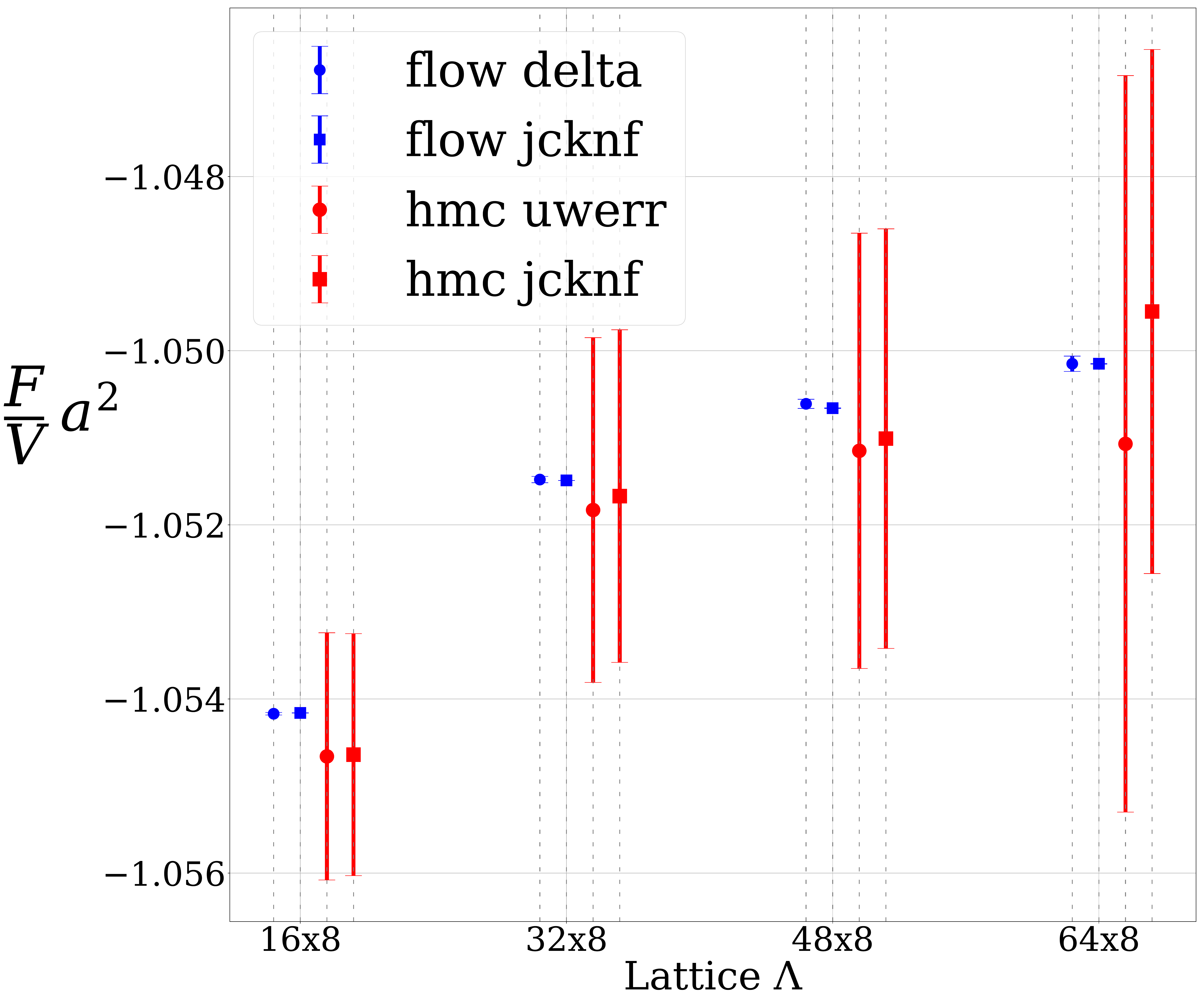}
    \caption{Free energy estimation with error analysis by both Jackknife and delta method. Both lead to compatible results. We use the same data as in Figure~\ref{fig:freeeneergy}.}
    \label{fig:freeenergyapp}
\end{figure}

\paragraph{Error estimation:} we use both the uwerr \cite{wolff2004monte} and jackknife method to estimate uncertainties for HMC. In order to deal with autocorrelation for jackknife, we perform binning with a 1k bin size. Error estimation for flow is performed by the delta method and also by jackknife, see Figure \ref{fig:freeenergyapp}. 

\begin{figure}[ht]
    \centering
    \vspace{0.8cm}
    \includegraphics[width=0.4\textwidth]{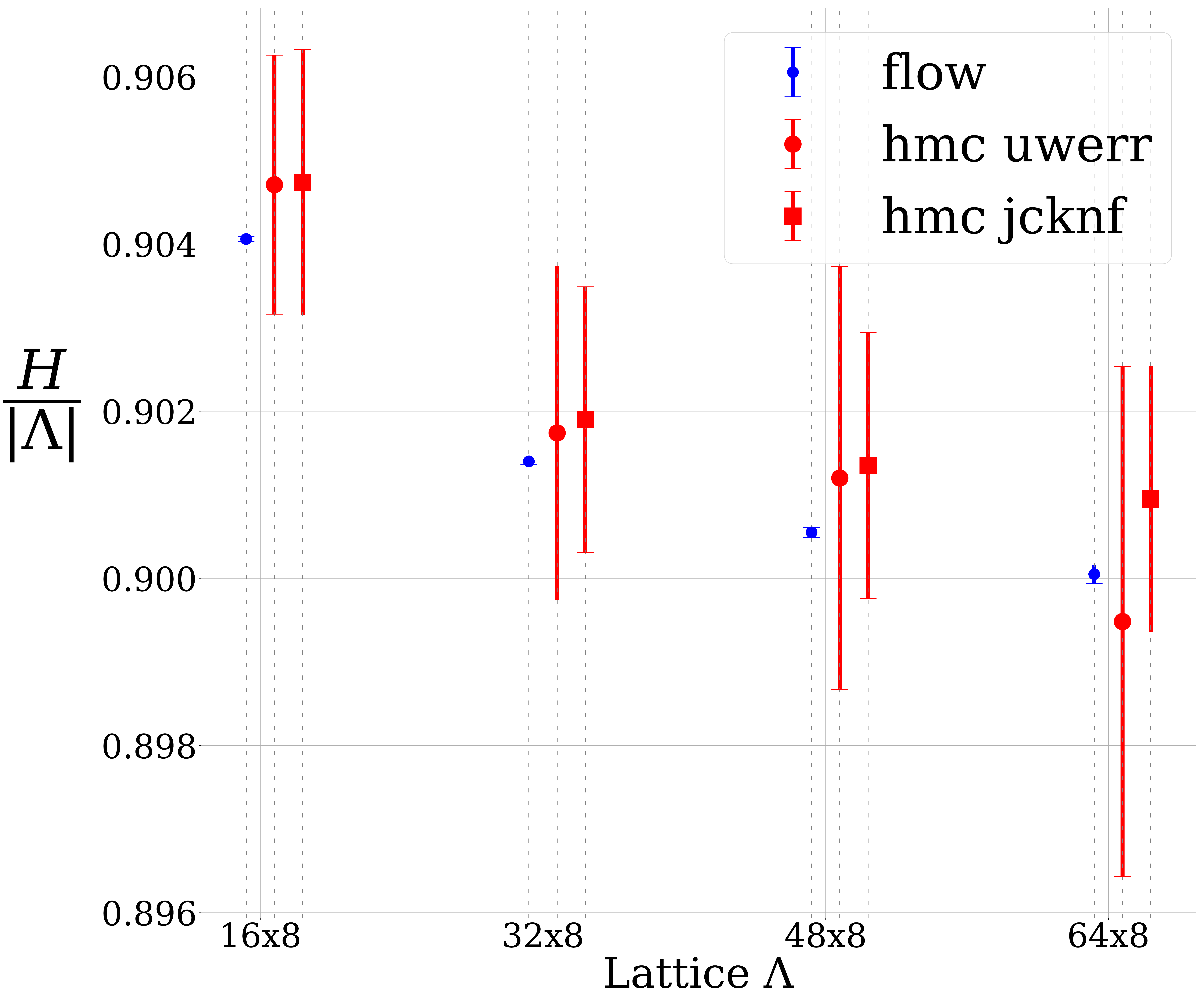}
    \caption{Entropy density estimation with error analysis by both Jackknife and delta method. Both lead to compatible results. We use the same setup as for Figure~\ref{fig:freeeneergy}.}
    \label{fig:entropyapp}
\end{figure}

From the free energy estimates, one can then derive other thermodynamic observables such as the entropy. We refer to the main text for a discussion of this. Figure \ref{fig:entropyapp} shows estimation of entropy. Errors were estimated using both the Jackknife and uwerr method. Both error analysis methods lead to consistent results. 

\paragraph{Runtime:}
 Normalizing Flows allow for very efficient sampling. Specifically, we can sample the total of 5.6M samples in under a minute for all lattices considered in the main text. Furthermore, this sampling procedure can be perfectly parallelized over multiple GPUs as the flow can generate each configuration independently. On the other hand, flows require a substantial up-front training cost which is independent of the number of samples used for free energy estimation and is therefore amortized over the entire estimation procedure. As a result, the relative runtime comparison between the flow-based and HMC-based algorithm strongly depends on the number of samples (as well as, of course, on the used implementation and hardware setup). In our numerical experiments, we use a Intel Xenon 2.4 GHZ CPU with NVidia P100 graphics card with 16GB memory. Both our implementation for the HMC as well as for the generative model were not optimized. The training time of the generative models takes about 20 hours to converge. Generating the samples with 14 different Markov chains takes about 25 hours. It is likely that both the MCMC runtime and training time could be substantially reduced by using more efficient implementations and by tuning the hyperparameters of the training process, such as initial learning rate, its decay schedule, choice of optimizer, initialization of weights, as well as early stopping (see \cite{montavon2012neural} for an overview). Since the goal of this work was to introduce the method and provide a proof of principle, we did not explore such techniques as part of this study.  

\end{document}